\documentclass[fleqn,usenatbib]{mnras}

\usepackage[pdftex]{graphicx}
\usepackage{epstopdf}
\usepackage{threeparttable}
\usepackage{newtxtext,newtxmath}
\usepackage{colortbl}
\usepackage[T1]{fontenc}
\usepackage{multicol}
\usepackage[caption=false]{subfig}

\DeclareRobustCommand{\VAN}[3]{#2}
\let\VANthebibliography\thebibliography
\def\thebibliography{\DeclareRobustCommand{\VAN}[3]{##3}\VANthebibliography}

\usepackage{graphicx}	
\usepackage{amsmath}	
\renewcommand{\d}[1]{\ensuremath{\operatorname{d}\!{#1}}}

\usepackage{xcolor}

\title[Gravity wave dissipation in star-planet systems]{Observational imprints of tidal internal gravity wave dissipation in star-planet systems}

\author[Lazovik \& Barker]{
Yaroslav A. Lazovik$^{1}$\thanks{E-mail: yaroslav.lazovik@gmail.com} and 
Adrian J. Barker$^{2}$\thanks{E-mail: A.J.Barker@leeds.ac.uk} 
\\
$^{1}$Sternberg Astronomical Institute, Lomonosov Moscow State University, Universitetsky pr. 13, Moscow, 119234, Russia \\
$^{2}$ School of Mathematics, University of Leeds, Leeds LS2 9JT, UK
}

\date{Accepted XXX. Received YYY; in original form ZZZ}

\pubyear{\the\year{}}

\begin{document}

\maketitle

\label{firstpage}
\pagerange{\pageref{firstpage}--\pageref{lastpage}}
\begin{abstract}
Tidal interactions play a crucial role in the orbital evolution of close-in star-planet systems. There are numerous manifestations of tides, including planetary orbital migration, breaking resonant chains, tidal heating, orbital circularization, spin-orbit alignment, and stellar and planetary spin synchronization. In the present study, we focus on the dissipation of internal gravity waves within stars. We examine two mechanisms: wave breaking in stars with radiative cores and magnetic wave conversion in stars with convective cores. Applying tidal prescriptions modelling these processes, we demonstrate that the enhanced stellar rotation of both TOI-2458 and GJ 504 can be explained by the previous engulfment of a hot Jupiter caused by gravity wave damping. Furthermore, we show that the observed population of hot Jupiters can be divided into two distinct subsamples: those that are too young for gravity wave dissipation and those where it is ongoing. These subsamples exhibit qualitatively different orbital period distributions: young systems have a uniform distribution, while older systems show a steep decline at short orbital periods. Using a population synthesis approach, we successfully reproduce the main features of the older hot Jupiter sample based on the distribution of the younger systems. According to our estimates, up to 13\% of the main-sequence stars within the mass range [0.7,1.5] $M_{\odot}$ that once hosted a hot Jupiter may have since engulfed it. Our results highlight the key role of internal gravity wave dissipation in shaping the orbital architectures of hot Jupiter systems.
\end{abstract} 




\section{Introduction}

Gravitational tidal interactions play an important role in driving both spin and orbital evolution in planetary systems and binary stars. The tidal gravity of one body (such as a star or planet) deforms and excites time-dependent flows inside another, and the dissipation of these flows results in tidal torques and spin-orbit evolution. In planetary systems, tidal dissipation inside stars is thought to cause planetary orbital migration for the closest-in planets, and in particular to be responsible for shrinking the semi-major axes of planets orbiting with periods shorter than the stellar rotation \citep[e.g.][]{Rasio1996,Jackson2009,L21}. Young stars that rotate more rapidly than their planet orbits them can potentially drive outward migration \citep[e.g.][]{BO2009,Bolmont2016,L21}. Tidal dissipation inside stars and planets can also modify the eccentricities of planetary orbits, typically driving orbital circularisation for hot Jupiters (HJs) \citep[e.g.][]{Jackson2008,Hansen2010,Mahmud2023,L2024}. It is also believed to modify stellar and planetary rotations \citep[e.g.][]{Maxted2015,Penev2018,Tejada2021,Ilic2024} and stellar and planetary obliquities \citep{BO2009,Albrecht2012,LO2017,Zanazzi2024}. 

The mechanisms of tidal dissipation in stars and planets are incompletely understood, though much progress has been made in recent decades \citep[e.g.][]{Ogilvie2014,Mathis2019,B2025}. 
Various attempts have been made to constrain efficiencies of tidal dissipation in stars from the observed populations of HJs around main-sequence \citep{Jackson2008,Jackson2009,Hansen2010,Hansen2012,CCJ2018,HamerSch2019,HamerSch2020,Miyazaki2023,Chen2023,Banerjee2024,Milholland2025} and subgiant stars \citep{SchWinn2013}. Other studies have constrained these efficiencies from the inferred stellar spin-up of main-sequence (MS) stars that would have been expected to rotate more slowly in the absence of tides \citep{Maxted2015,Penev2018,Ilic2024}. There have also been attempts to constrain the efficiencies of tidal dissipation in binary stars \citep[e.g.][]{Penev2022} and giant planets \citep[e.g.][]{Jackson2008,Mahmud2023,L2024}.

Theoretically, various mechanisms of tidal dissipation have been proposed to operate in low-mass planet-hosting stars. Classically, convective damping of large-scale equilibrium tides has been hypothesized to be important \citep[e.g.][]{Z1977,GN1977,Z1989}, but recent hydrodynamical simulations have indicated that this mechanism is unlikely to be effective in MS stars \citep[in the so-called ``fast tides'' regime, which is likely to be relevant for the energetically dominant convective eddies e.g.][]{Penev2009,OL2012,DBJ2020b,VB2020b,T2021,BA2021,Nils2023}. Instead, the tidal excitation and dissipation of various kinds of waves, including inertial waves in convection zones of sufficiently rapidly rotating stellar hosts \citep{OL2007,Bolmont2016,Barker,Wu2024}, and internal gravity waves (IGWs, or g-modes, or inertia-gravity/gravito-inertial waves when rotation is non-negligible) in radiative zones \citep{Goodman,T1998,OL2007,BO21,IvPap2013,Barker,Ahuir21,Guo23,Esseldeurs2024}. Resonance locking of the latter has also been proposed \citep{MaFuller2021}, which involves tidal and g-mode frequencies evolving in such a way that they can remain locked in resonance with the tidal forcing, leading to enhanced dissipation \citep[as first proposed and studied by][]{Witte1999}. 

Recently, tidal dissipation mechanisms are beginning to be constrained via transit timing analysis of HJs \citep[e.g.][]{Birkby2014,Maciej2016,Wilkins2017,Patra2020,Yee2020,Vissap2022,B24,Adams2024,Basturk2025}. The idea is that the gradual orbital evolution of short-period planets can be detected by shifts in the arrival times of HJs, potentially allowing us to constrain stellar tidal dissipation if other processes can be excluded from causing any observed changes. Such more direct detections of stellar tidal dissipation complement statistical and population-wide approaches to constraining tidal mechanisms and their efficiencies in stars (and planets). In this study, both the analysis of individual systems and a study of the population of HJs as a whole are of interest to us.

Tidal interactions have been explored extensively in studies dedicated to the evolution of the HJ population. Based on an equilibrium tide model, \cite{CCJ2018} attempted to explain the upper-left boundary in the mass–separation diagram. Similar results were obtained by \cite{Rao2021}, whose model incorporates both equilibrium tide and inertial wave dissipation. Using the same tidal prescriptions, \cite{Bolmont2017} earlier examined the influence of metallicity on the populations of HJs orbiting stars with different initial spin rates. Inertial waves were also utilized by \cite{Heller2019} to investigate the formation of the HJ pile-up at approximately 0.05 AU. Applying crude equilibrium tide and stellar wind models, \cite{Ferraz-Mello_Beauge2023} successfully reproduced the distribution of HJ systems in the orbital period–rotation period diagram. Furthermore, \cite{Ahuir21} employed tidal and magnetic star–planet interactions within their synthetic populations to reproduce the observed rotation period distribution of HJ hosts and the orbital period distribution of HJs. Beyond these studies, some aspects of tidal interactions have also been integrated into global population synthesis models designed to trace the formation and subsequent evolution of planets, including HJs \citep{Emsenhuber2021}.

Despite the aforementioned advancements, IGWs have yet to be thoroughly explored in the context of star-planet orbital evolution and its effects on the planetary population until recently. In \cite{L21, L23}, we demonstrated that IGW dissipation via strongly nonlinear wave breaking can significantly influence the orbital architectures of HJ systems. However, our population synthesis analysis was restricted by underlying assumptions regarding the initial hot Jupiter population. In the present work, we aim to address these limitations. 

Our primary focus is on two key mechanisms: wave breaking in stars with radiative cores \citep[e.g.][]{BO21} and magnetic wave conversion \citep[e.g.][]{Duguid24} in stars with convective cores. These processes are expected to provide rapid dissipation rates during the second half of the host stars' main-sequence lifetimes, creating a stark contrast between the orbital period distributions of planets in younger and older systems. The paper is organized as follows. Sec. \ref{sec:methods} details our tidal prescriptions and planetary migration model. In Sec. \ref{sec:spin-up}, we use this model to investigate tidally-driven engulfment as a potential cause for the enhanced rotation rates of TOI-2458 and GJ 504. In Sec. \ref{sec:population}, we conduct a population synthesis analysis to assess the effect of IGW dissipation on the evolution of observed HJs. Our results are discussed in Sec. \ref{sec:discussion} and summarised in Sec. \ref{sec:conclusions}.

\section{Methods}
\label{sec:methods} 
In this section, we present the theoretical framework used to model planetary migration due to IGW dissipation. In tidal theory, a fundamental quantity for modelling tidal dissipation is the stellar modified tidal quality factor, $Q'$. It is defined as being proportional to the ratio of the maximum potential energy stored in the tide ($E_0$) to the energy dissipated over one tidal period, according to:
\begin{equation}
    Q' = \frac{3}{2k_2} \frac{2 \pi E_0}{ \oint  \left(- \frac{\d E}{\d t} \d t \right)},  
	\label{eq:Qfactor}
\end{equation}
where $k_2$ is the second-order potential Love number (which is thought to take the value $k_2=0.0351$ for the Sun). Although this quantity is not a constant parameter, and in general it depends upon stellar mass, rotation, age, and metallicity, as well as on the tidal amplitude and frequency \citep[e.g.][]{Ogilvie2014,B2025}, it remains a helpful quantity because it appears in tidal evolutionary equations. In this work, we focus on circular planetary orbits in the equatorial plane of a star, in which the relevant tidal component is the quadrupolar one with spherical harmonic degree $l=2$ and azimuthal wavenumber $m=2$, which has tidal frequency $\omega_\mathrm{tide}=2(n-\Omega_\ast)$. Here $\Omega_{*}$ is the stellar rotation frequency and $n=\sqrt{G(M_{\textrm{pl}}+M_{\star})/a^3}$ is the orbital frequency (mean motion), where $M_{\textrm{pl}}$ is the planetary mass, $M_{\star}$ is the stellar mass, $G$ is the gravitational constant and $a$ is the orbital semi-major axis.

\subsection{Dissipation mechanisms}
\label{subsec:mechanisms}

In the present study, we focus on two processes responsible for the dissipation of tidally excited IGWs: wave breaking and conversion to magnetic waves. These are thought to be relevant for stars with radiative and convective cores, respectively. They are expected to provide the most efficient dissipation of IGWs when they operate unless tidal forcing can resonantly excite a global g-mode oscillation for a sustained period of time, in which case it is in principle possible for the dissipation to be stronger \citep[at least in linear theory, e.g.][]{T1998,Goodman,Witte1999,MaFuller2021}. However, nonlinear effects and evolution of the internal rotation of a star may prevent resonances from being maintained or locked into in many cases \citep[e.g.][]{BO21,MaFuller2021,Guo23}. Hence, the two mechanisms above are likely to be the most efficient ones because they lead to the deposition and subsequent dissipation of all of the energy stored in the tidal waves inside the star after their excitation. The corresponding tidal quality factor $Q'$, characterizing the efficiency of tidal dissipation in this ``fully damped regime'', in which IGWs are launched from the boundary between the convective envelope and radiation zone as ``travelling waves'', can be expressed as \citep[e.g.][]{Barker}:
\begin{equation}
    \frac{1}{Q'} = \frac{2 [\Gamma(\frac{1}{3})]^2}{3^{\frac{1}{3}}(2l+1)(l(l+1))^{\frac{4}{3}}} \frac{R_{*}}{G M_{*}^2} \mathcal{G} |\omega_\mathrm{tide}|^{\frac{8}{3}}.
    \label{eq:tide_gw1}
\end{equation}
The parameter $\mathcal{G}$ is calculated following equation~(42) of \cite{Barker}. 

While equation~(\ref{eq:tide_gw1}) in principle applies if the waves are fully damped according to any mechanism (such as radiative diffusion of low frequency waves), we will apply it only when either the waves have large enough amplitudes to become strongly nonlinear and break, or via the linear process of magnetic wave conversion. We do not consider additional linear and weakly non-linear processes explicitly, such as those involving weaker radiative damping of IGWs \citep[the neglect of which is more appropriate here than in more massive stars with radiative envelopes, e.g.][]{Z1975,Z1977,SavPap1983,GN1989} or global g-modes \citep{Goodman}, and we also ignore excitation of secondary waves by weakly nonlinear interactions \citep{Essick, Weinberg}, which have been explored in some works in the literature. Studying their impact on star-planet orbital evolution is left to future work. Furthermore, equation~(\ref{eq:tide_gw1}) assumes that the squared buoyancy frequency $N^2$ at the radiative/convective boundary depends linearly upon the distance from the boundary. As demonstrated by \cite{B11}, accounting for deviations from this linear profile may introduce a factor of $\sim 2$ variation in $Q'$. However, such differences would be insufficient to alter the primary results of this paper.

In order to isolate the implications of IGW dissipation from the action of inertial waves in convection zones, in Section~\ref{sec:population} we impose age restrictions, filtering out the young systems where the host star can be a rapid rotator and could be subject to efficient inertial wave dissipation. We also neglect the contribution from equilibrium tides, which are expected to be ineffective in all stars considered here due to the reduction in turbulent viscosity for fast tides \citep[e.g.][]{GN1977,OL2012,VB2020b,DBJ2020b}. We also do not take into account non-tidal mechanisms driving the planetary orbital migration, such as star-planet magnetic interactions \citep[e.g.][]{Lai,Strugarek16,Strugarek17,Ahuir21,WeiLin,Colle25}. The latter is justified during the later stages of MS evolution because the magnetic fields of both solar-type stars \citep{Ahuir20} and HJs \citep{Hori,Kilmetis} weaken with age, thereby reducing the significance of magnetic interactions.

As a result of our choice to neglect the processes just described, we assume that the planetary orbit is static until the conditions required for IGW dissipation, outlined below, are satisfied.

\subsubsection{Wave breaking in cooler stars with radiative cores}
\label{subsubsec:breaking} 
In stars with radiative cores, IGWs are excited at the interface between the radiative core and the convective envelope. While approaching the stellar centre, the amplitudes of these waves increase due to geometrical focussing, such that they may become non-linear there. Once a critical amplitude is exceeded, these waves overturn the background stratification, leading to wave breaking \citep{BO21}. We assume that wave breaking occurs if the following condition is satisfied \citep[see e.g.,][]{Barker}:
\begin{equation}
A_\mathrm{nl}^2 = \frac{3^{\frac{2}{3}}54\sqrt{6}[\Gamma(\frac{1}{3})]^2}{25\pi(l(l+1))^{\frac{4}{3}}} \frac{\mathcal{G} C^5}{\rho_0} \left(\frac{M_\mathrm{pl}}{M_{*}}\right)^2 \left(\frac{R_{*}}{a}\right)^6 |\omega_\mathrm{tide}|^{-\frac{13}{3}} \gtrsim 1,
    \label{eq:tide_gw2}
\end{equation}
with $\rho_0$ the central stellar density, $\mathcal{G}$ the stellar structural parameter relating to the radiative/convective interface described above, and $C$ the slope of the buoyancy frequency profile near the centre of the star (such that the buoyancy frequency $N=Cr$ there, where $r$ is the distance from the centre). $C(t)$ evolves with stellar age $t$, and increases substantially with time for ages older than 1 Gyr, e.g.~see Fig.~10 in \citet{Barker}. For a given system, the condition $A_\mathrm{nl}^2=1$ (in equation~\ref{eq:tide_gw2}) can then be used to define the age at which $t=t_\mathrm{gw}$, corresponding to the onset of IGW breaking. 

We also note that other processes responsible for IGW damping, such as viscous and thermal dissipation, or resonant excitation of g-modes leading to wave breaking for smaller amplitudes than predicted by equation~(\ref{eq:tide_gw2}), can modify stellar rotation and lead to the formation of a critical layer, where incoming waves are effectively absorbed \citep{BO21,Guo23}. Consequently, we expect that the condition (\ref{eq:tide_gw2}) defines a conservative condition for the efficient dissipation of IGWs.

\subsubsection{Conversion to magnetic waves in hotter stars with convective cores}
\label{subsubsec:conversion} 
IGWs cannot propagate within convective layers such as the convective cores of F-type stars. This prevents the waves from reaching the stellar centre in these stars, and hence IGWs are unlikely to achieve sufficiently large amplitudes to cause them to break when they are excited by planetary mass companions \citep[see Fig.~9 of][]{Barker}. However, wave breaking is not the only mechanism that can lead to efficient dissipation of IGWs. Based on observations of depressed dipole ($l = 2$) oscillation modes in red giant branch (RGB) stars, \cite{Fuller15} proposed that inwardly propagating IGWs may be converted into outwardly propagating magnetic (Alfvénic or slow magnetosonic) waves in regions where the radial wavenumbers of each of these waves become comparable. This picture is based on an unmagnetised or weakly magnetised envelope in which IGWs are excited and propagate inwards until they reach a region of sufficiently strong magnetic field where the above-mentioned condition is satisfied. These magnetic waves are likely to subsequently dissipate as they propagate outward into unmagnetized regions of the star, partly because they typically have much shorter wavelengths than the ingoing wave. This is therefore a mechanism for dissipating inwardly propagating IGWs. This process has been further investigated in \cite{Lecoanet17} and \cite{RuiFuller23}, which generally support the suggestion of \cite{Fuller15} (though see \citealt{Loi2017,Loi2018}).

The impact of IGW conversion into magnetic waves on tidal dissipation and for the orbital evolution of star-planet systems has begun to be studied in \cite{Duguid24}, where its potentially significant role for HJs and ultra-short period planets (USPs) was highlighted. They found that the conditions necessary for wave conversion to occur are typically met in stars with convective cores during the second half of the main sequence when the convective core retreats. This allows a strong magnetic field to exist in their inner radiative layers, which was originally produced by a convective core dynamo prior to the core retreating. Following \cite{Duguid24}, we assume that IGWs are fully converted into magnetic waves when the radial magnetic field strength, with magnitude $|B_r|$, exceeds the critical value $B_\mathrm{crit}$, at any point within the region $R_\mathrm{c} < r < R_\mathrm{c} + 0.2 R_{\star}$, where
\begin{equation}
B_r(t) = \frac{6\sqrt{\pi}}{R_\mathrm{c}^3}\int_0^{R_\mathrm{c}}r^2\sqrt{\rho u_\mathrm{conv}^2} \rm dr,
    \label{eq:magn1}
\end{equation}
and
\begin{equation}
B_\mathrm{crit}(r,t) = \frac{\pi \omega_\mathrm{tide}^2 r \sqrt{\rho}}{N}.
    \label{eq:magn2}
\end{equation}
Here, $R_\mathrm{c}$ is the radius of the convective core, $\rho(r)$ is the local density, $u_\mathrm{conv}(r)$ is the convective velocity, and $N(r)$ is the Brunt–Väisälä (buoyancy) frequency (the latter three depend upon the distance from the stellar centre, $r$). By imposing the constraint $R_\mathrm{c} < r < R_\mathrm{c} + 0.2 R_{\star}$, we aim to exclude the upper layers of the radiative envelope, where the global minimum of $B_\mathrm{crit}$ is located during the early stages of MS evolution \citep[see][]{Duguid24}, and to focus on the core-envelope boundary. The condition $|B_r| \geq B_\mathrm{crit}$ determines the onset of IGW dissipation in stars possessing convective cores in our models. Unlike the wave-breaking criterion discussed in subsection~\ref{subsubsec:breaking}, this condition does not depend on the planetary mass (except through its very weak influence on $\omega_\mathrm{tide}$).

As a planet migrates inward, its orbital period decreases (hence $\omega_\mathrm{tide}$ increases). This decrease can cause the critical magnetic field strength to rise above $B_r$ in cases where magnetic wave conversion has previously occurred. In these cases, we assume that inwardly propagating IGWs continue to be fully dissipated after magnetic wave conversion has started, and hence the resulting tidal dissipation is determined by equation~(\ref{eq:tide_gw1}) even after the criterion $B_r>B_\mathrm{crit}$ is no longer satisfied. Currently, there are many unsolved problems relating to the propagation and dissipation of IGWs in regions with strong magnetic fields, so the validity of this approach is ultimately uncertain, though we come back to this point later in Section~\ref{subsec:uncert}.

Finally, we note that although there are two separate underlying processes responsible for IGW dissipation in stars with radiative and convective cores, here we use the same notation $t_\mathrm{gw}$ to represent the age at which efficient IGW dissipation (governed by equation~\ref{eq:tide_gw1}) begins. The mechanism involved that leads to this regime is determined by the stellar type, particularly whether it has a convective or radiative core.
  
\subsection{Orbital evolution}
\label{subsec:evolution} 
In the present paper, we assume that by the time efficient dissipation of IGWs onsets, the planetary orbit is circular and aligned with the stellar equatorial plane. The evolution of the semi-major axis is determined by:
\begin{equation}
    \frac{1}{a}\frac{\d a}{\d t} = \frac{\Omega_{*} - n}{|\Omega_{*} - n|}\frac{9n}{2} \left(\frac{M_\mathrm{pl}}{M_{*}} \right) \left(\frac{R_{*}}{a} \right)^5 \frac{1}{Q'}.
	\label{eq:orbit7}
\end{equation} 
Hereafter, we assume $M_{\star} \gg M_\mathrm{pl}$ as $M_\mathrm{pl}/M_{\star} < 2 \times 10^{-4}$ in our work.

The transfer of angular momentum between the host star and the planetary orbit also changes the stellar rotation rate. We consider magnetic braking, quantified by wind torque $\Gamma_\mathrm{wind}$, to be the only source of total angular momentum loss from the star-planet system (note that $\Gamma_\mathrm{wind}$ is negative in the present study). Accordingly, we have: 
\begin{equation}
    \frac{\d L_{*}}{\d t} + \frac{\d L_\mathrm{pl}}{\d t} = \Gamma_\mathrm{wind},
	\label{eq:orbit1}
\end{equation}
where $L_{*}=  I_{*}\Omega_{*}$ and $L_\mathrm{pl}=\,M_\mathrm{pl}\, a^2 n$ are the stellar spin and planetary orbital angular momenta, respectively, and $I_{\star}$ is the stellar moment of inertia \citep[computed as described in e.g.][]{Barker,L21}. Accordingly, the evolution of the stellar rotation rate can be computed by the following equation: 
 \begin{equation}
    \frac{\d \Omega_{*}}{\d t} = \frac{1}{I_{*}} \left(\Gamma_\mathrm{wind} - \Omega_{*} \frac{\d I_{*}}{\d t} - \frac{1}{2} M_\mathrm{pl} \sqrt{\frac{G M_{*}}{a}} \frac{\d a}{\d t} \right).   
	\label{eq:orbit2}
\end{equation}
As in \cite{L21,L23}, we implement the braking law by \citet{Matt} and \citet{Amard} with the coefficients parametrizing the wind torque, $\Gamma_\mathrm{wind}$, adopted from \cite{Gossage}. 

When examining the impact of planetary engulfment on stellar rotation in Section~\ref{sec:spin-up}, we assume that the planet instantaneously transfers all of its orbital momentum to the star's rotation once its semi-major axis equals the Roche limit $a_\mathrm{R}$, which is evaluated as follows:
\begin{equation}
a_\mathrm{R} = f_\mathrm{p} R_\mathrm{pl} \left({\frac{M_\mathrm{*}}{M_\mathrm{pl}}}\right)^{\frac{1}{3}}. 
	\label{eq:roche}
\end{equation}
The parameter $f_\mathrm{p}$ depends on the planetary structure and constitution. In the present study, $f_\mathrm{p}$ is fixed at 2.4, which is approximately valid for a homogeneous fluid body. This value is higher than the value $f_\mathrm{p} = 2.16$ in e.g.~\cite{Faber2005,Ford2006,Nelson2017}, but lower than the value $f_\mathrm{p} = 2.7$ inferred from the three-dimensional hydrodynamical simulations of \citet{Guillochon2011}. Although this work assumes tidal disruption upon filling the Roche lobe, another possibility is the stable accretion scenario proposed by \citet{Valsecchi} and further investigated by \citet{L23}. According to this scenario, HJs can avoid immediate tidal disruption by continuously adjusting their radii to fill their Roche lobes while gradually losing mass, eventually transforming into hot Neptunes with thin envelopes, or Super-Earths. However, the validity of this scenario relies on an efficient mechanism to return the angular momentum of the transferred mass back to the planet's orbit. The existence of such a mechanism was questioned by \citet{Jia}, who argued that the star acts as a sink for the angular momentum, preventing orbital expansion and instead accelerating runaway mass transfer, which in the end causes tidal disruption. Recently, \citet{Hallatt} proposed an impulsive mass-loss regime, according to which a small return of angular momentum ($\sim 5-10\%$) is able to prevent engulfment, leading to the rapid shedding of most of the planet's envelope. Nevertheless, even if a planet can survive the mass loss following Roche lobe overflow, we expect IGW dissipation to be sufficiently strong to subsequently drive inspiral into the star within a short timescale.

Throughout our work, we assume that the star rotates as a solid body. Hence we neglect any possible differential rotation (or ``core-envelope decoupling") between convection and radiation zones, and also within convective or radiative regions. It is important to note that the considered mechanisms of IGW dissipation will alter the rotation profile preferentially within the radiation zone where these waves are damped, and hence where the tidal torque will apply, which is most likely to occur deep within the star. At the same time, planetary engulfment, occurring at the final stage of tidal decay, primarily spins up the outermost stellar layers. The timescales required to transport angular momentum between and within stellar zones (both radiative and convective) are currently very uncertain. In their work, \cite{GalletBouvier2015} constrained core-envelope coupling timescales to the range $10^7-10^8$ yrs to match the observed distribution of stellar rotation at ZAMS, which is shorter than most of the timescales relevant to this study. Nevertheless, exploring the effects of differential rotation and its coupled interaction with tidal dissipation is an important topic for future studies.

\subsection{Stellar models}
\label{subsec:models} 

\begin{figure}
	\includegraphics[width=\columnwidth]{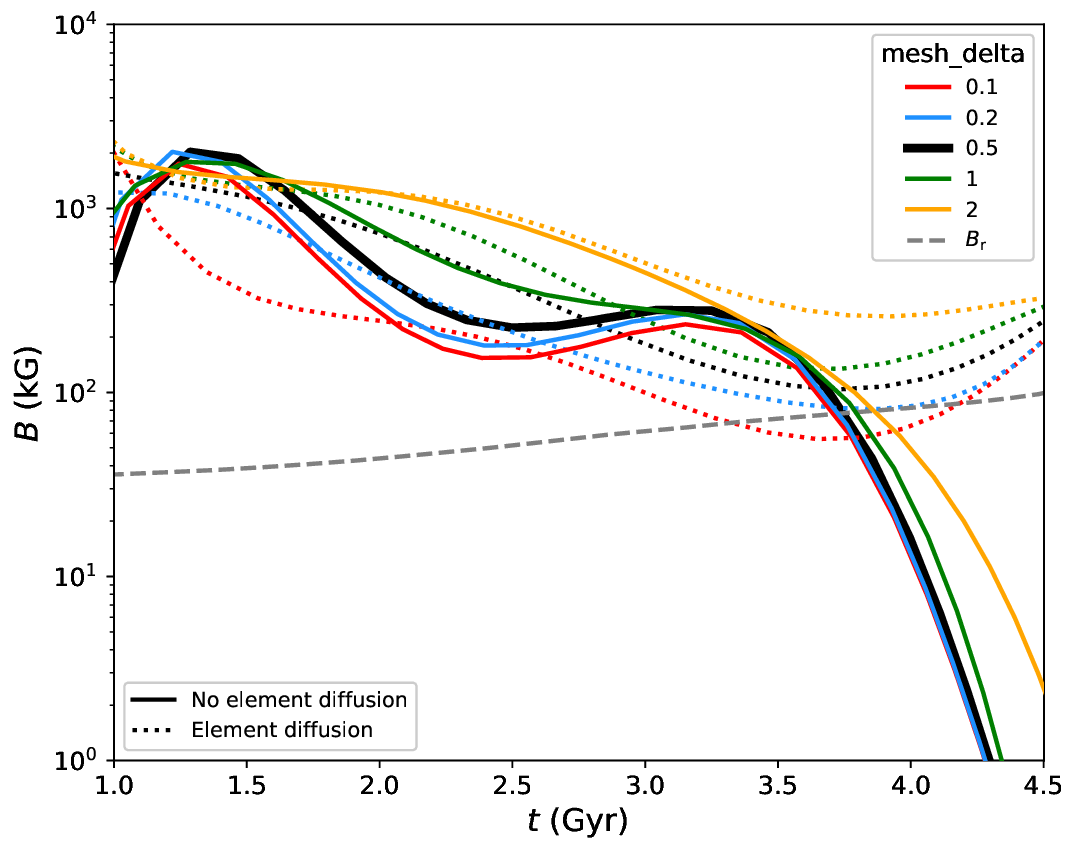}
    \caption{Evolution of the critical magnetic field strength, $B_\mathrm{crit}$, for a 1.2 $M_{\odot}$ solar-metallicity model and tidal period of 0.5 days. Each color represents a different grid resolution, defined by the parameter \texttt{mesh\_delta\_coeff}, given in the legend. $B_\mathrm{crit}$ is evaluated as a local minimum within the region [$R_{\mathrm{c}}, R_{\mathrm{c}} + 0.2 R_{*}$]}. Solid (dotted) lines correspond to models without (with) element diffusion. Gray dashed line shows the radial magnetic field strength of the convective core, $B_r$.
    \label{fig1}
\end{figure}

In this study, we compute our stellar models using the evolution code \textsc{MESA} r11701 \citep{MESA1,MESA2,MESA3,MESA4,MESA5} and inlist files from \cite{Gossage} based on the MIST framework \citep{Dotter,Choi}. The models account for stellar rotation, parameterized according to the magnetic braking law of \citet{Matt, Amard}. The initial rotation period following disc dissipation is set to 5.5 days, consistent with typical rotation rates observed in NGC 2362 \citep{Irwin}. The disc dissipation timescale is fixed at 6 Myr, based on \citet{Tu}. Since the star-planet systems explored in this paper are older than 100 Myr, their present-day rotation rates are insensitive to the initial conditions; the same is true for rotation rates at the onset of IGW dissipation. We find that the critical magnetic field strength required for wave conversion, $B_\mathrm{crit}$, is sensitive to the grid resolution, which is determined by the parameter \texttt{mesh\_delta\_coeff}. Increasing the resolution (by decreasing \texttt{mesh\_delta\_coeff}) allows $B_\mathrm{crit}$ to converge at \texttt{mesh\_delta\_coeff} $\lesssim 0.5$, as we show in Fig.\ref{fig1}. Therefore, we select \texttt{mesh\_delta\_coeff} $ = 0.5$ to balance computational efficiency and accuracy.

Furthermore, Fig.\ref{fig1} demonstrates that the evolution of $B_\mathrm{crit}$ is qualitatively and quantitatively different depending on whether element diffusion is considered. In particular, implementing diffusion eliminates the bump in the $N^2$ profile near the inner edge of the radiative layer, which is present in models computed without element diffusion at late ages \citep[see e.g.~Fig.~2 in][]{Duguid24}. This bump causes a drop in the evolution of $B_\mathrm{crit}$ and thus enables the conversion of IGWs over a wide range of tidal forcing periods. 
We note that current understanding of the efficiency of element diffusion and its impact on the evolution at the interface between the convective core and the radiative envelope remains uncertain. Additionally, calculating $B_r$ based on the assumption of equipartition between magnetic and convective kinetic energy densities may lead to an underestimate of the radial magnetic field strength \citep[compared to, for example, the magnetostrophic scaling as in Table 1 of][]{Astoul2019}. Given these considerations, we have decided to disable element diffusion for the remainder of this work to ensure that our estimates are consistent with the models considered by \cite{Duguid24}. However, this issue remains a substantial uncertainty in our models. 

\section{Stellar spin-up as a result of planetary engulfment}

Before examining how IGW dissipation affects the overall HJ population, we will focus on two particular cases where tidally driven migration leading to planetary engulfment of a hypothetical HJ can plausibly explain the observed increase in stellar rotation rates compared with expectations for single stars. Specifically, we will study the stars TOI-2458 and GJ 504, which have either a radiative or a convective core, respectively. We will demonstrate that IGW dissipation is a plausible mechanism capable of driving the migration of a hypothetical HJ at later ages, ultimately causing planetary engulfment. Importantly, our analysis will enable us to accurately reproduce the current ages and rotation rates of these stars without the need to finely tune the parameters of the engulfed planets. This supports IGW dissipation as a key process in the evolution of close-in exoplanetary systems and their host stars.

\label{sec:spin-up} 
\subsection{TOI-2458}

\begin{figure}
	\includegraphics[width=\columnwidth]{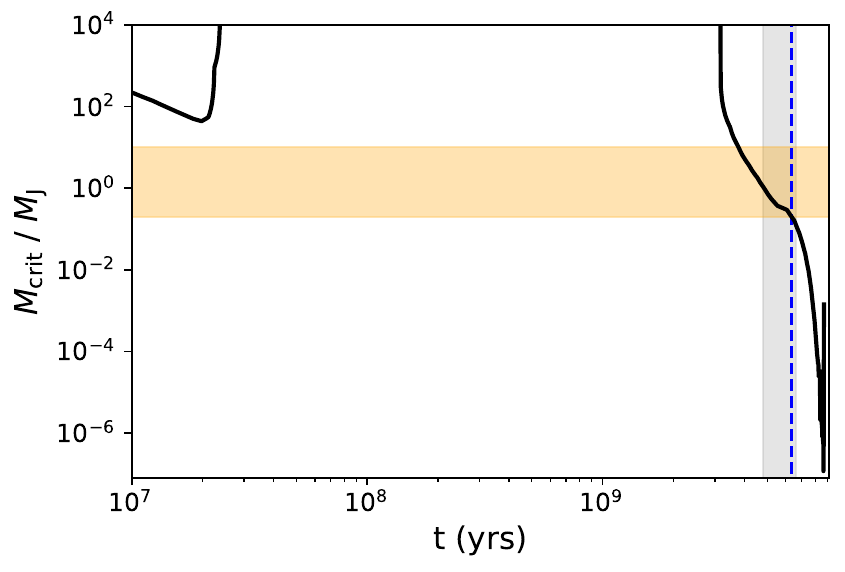}
    \caption{Critical planetary mass for wave breaking as a function of age for TOI-2458 and tidal period of 0.5 days. Stellar mass and metallicity are $M_{*}$ = 1.05 $M_{\odot}$, [Fe/H] = -0.11. Blue vertical dashed line illustrates the TAMS age. Gray region indicates the age constraint from \protect\cite{Subjak}. Orange region indicates HJ mass range.}
    \label{fig2}
\end{figure}

TOI-2458 is a 1.05 $M_{\odot}$ main-sequence star with sub-solar metallicity ([Fe/H] = -0.11), hosting a hot Neptune, TOI-2458 b, with an orbital period $P_\mathrm{orb} =$ 3.74 days and a mass $M_\mathrm{pl}=13.3 M_{\earth}$ \citep{Subjak}. The rapid rotation rate ($P_\mathrm{rot}$ = 8.9 days; $P_\mathrm{rot}$ is the rotation period, whereas we would expect $P_\mathrm{rot}\approx 20$ days without tidal spin-up) for its age of $~5.7^{+0.9}_{-0.8}$ Gyr suggests that the host star may have been spun up due to tidal interactions with another massive planet before it was engulfed by the star. Given the presence of the close-in planets TOI-2458 b and c, the hypothetical planet that was engulfed most likely formed via disk migration -- or potentially in situ -- rather than by a more violent high-eccentricity migration scenario. The estimated age of TOI-2458 is close to the terminal-age main-sequence (TAMS). This allows wave breaking to operate for a broad range of planetary masses, as we show in Fig.~\ref{fig2}, which shows the critical planetary mass required to initiate wave breaking as a function of age in the core of TOI-2458, assuming a tidal period of $0.5$ days (approximately valid for a hypothetical planet with $P_\mathrm{orb}\sim 1$ d).

\begin{figure}
	\includegraphics[width=\columnwidth]{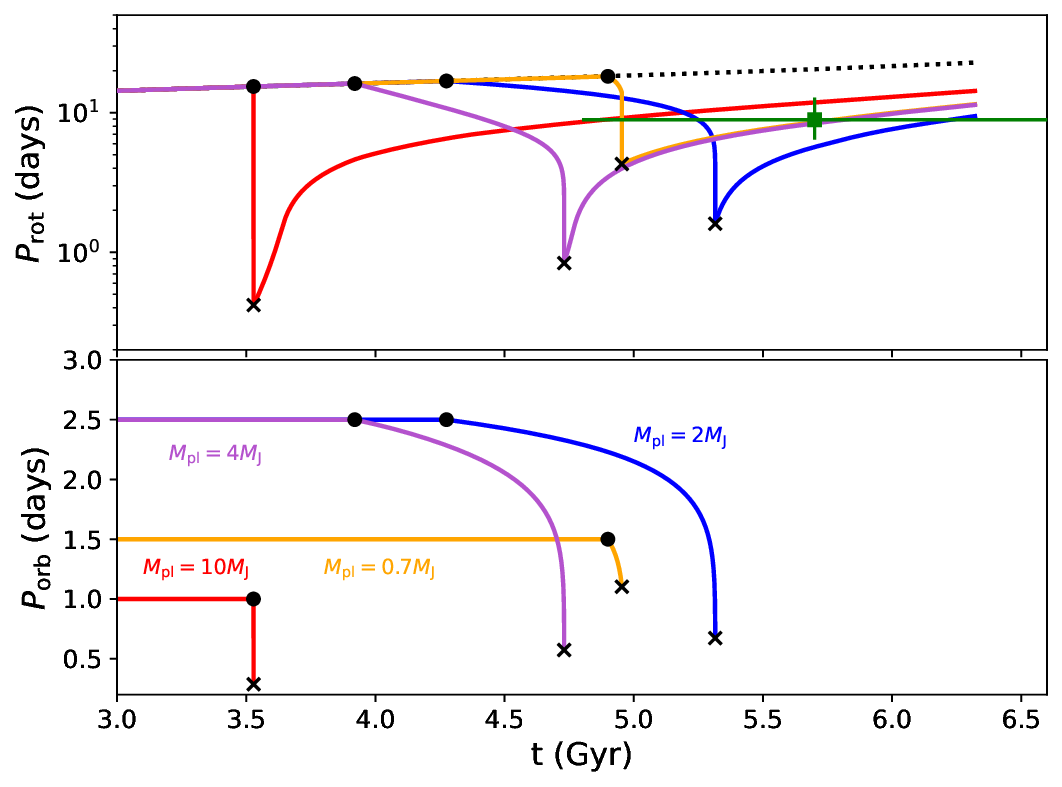}
\caption{Evolution of stellar rotation (top panel) and planetary orbital (bottom panel) periods for four star-planet systems with $P_\mathrm{orb}$ = 1, 1.5, 2.5, 2.5 days and $M_\mathrm{pl} = $ 10, 0.7, 2, and 4 $M_\mathrm{J}$, shown in red, orange, blue, and purple, respectively. The black dotted line depicts the rotation period evolution of a solitary star. Circles and crosses denote the onset of IGW breaking and Roche-lobe overflow (and planetary engulfment), respectively. The green square with error bars shows TOI-2458's observed rotation period and age, along with their uncertainties from \protect\cite{Subjak}.}
\label{fig3}
\end{figure}

Our hypothesis that IGW dissipation can explain the star's anomalously fast rotation is supported by Fig.~\ref{fig3}, illustrating the orbital evolution of four hypothetical HJs around TOI-2458, with planetary masses of 0.7, 2, 4, and 10 Jovian masses, and initial orbital periods of 1.5, 2.5, 2.5, and 1 day, respectively. These initial orbital periods represent the range of possible separations between the engulfed planet and its host star, considering the proximity of the hot Neptune TOI-2458 b. For the least massive HJs, such as those with masses as small as $0.3 \: M_J$, the smaller initial separations above would place them below the Roche limit (for $M_\mathrm{pl} = 0.3 \: M_\mathrm{J}$ and $R_\mathrm{pl} = 1.5 \: R_\mathrm{J}$, equation~(\ref{eq:roche}) yields $a_\mathrm{R} \approx 0.026$ AU, which corresponds to an orbital period of $\approx 1.5$ days). Conversely, a planet with a larger initial separation would be too close to TOI-2458 b, as the mutual separation between the two planets would fall below 4 mutual Hill radii, which is the stability threshold for two-planet systems proposed by \cite{Wu2019}. The mutual Hill radius $\Delta_\mathrm{H}$ in the units of mutual planetary separation is determined by:
  \begin{align}
    \frac{\Delta_\mathrm{H}}{\Delta a} = &\frac{a_1 + a_2}{2(a_2 - a_1)}\left(\frac{M_{\mathrm{pl},1} + M_{\mathrm{pl},2}}{3M_{\star}}\right)^{\frac{1}{3}} \nonumber\\& \approx 3.4 \times 10^{-2} \frac{P_\mathrm{orb,1}^{\frac{2}{3}}+P_\mathrm{orb,2}^{\frac{2}{3}}}{P_\mathrm{orb,2}^{\frac{2}{3}}-P_\mathrm{orb,1}^{\frac{2}{3}}} \left(\frac{M_\mathrm{pl,1} + M_\mathrm{pl,2}}{1 M_\mathrm{J}}\right)^{\frac{1}{3}}\left(\frac{M_\mathrm{\star}}{ M_\mathrm{\odot}}\right)^{-\frac{1}{3}},   
	\label{eq:hill}
\end{align}
 where $a_1$ and $a_2$ ($P_\mathrm{orb,1}$ and $P_\mathrm{orb,2}$) are the semi-major axes (orbital periods) of neighboring planets with masses $M_{\mathrm{pl},1}$ and $M_{\mathrm{pl},2}$, respectively. For $P_\mathrm{orb,1} = 2.5$ days, $M_\mathrm{pl,1} = M_\mathrm{J}$, $P_\mathrm{orb,2} = 3.74$ days, $M_\mathrm{pl,2} = 0.042 \; M_\mathrm{J}$, and  $M_\mathrm{\star} = 1.05 \; M_\mathrm{\odot}$ (the latter three are the parameters of TOI-2458 b and TOI-2458, respectively), equation~(\ref{eq:hill}) gives $\Delta_\mathrm{H}/\Delta a \approx 1/4$.

Rapid orbital migration, triggered by IGW breaking (the onset of which is shown by the black circles), modifies the stellar rotation period compared to that of a solitary star (shown with a black dotted line), as depicted in the top panel of Fig.~\ref{fig3}. The bottom left panel reveals that the engulfment of planets with $M_\mathrm{pl}$ = 0.7 and 10 $ M_\mathrm{J}$ occurs almost instantaneously after the onset of wave breaking. In contrast, 2.0 and 4.0 $M_\mathrm{J}$ planets undergo a prolonged phase of orbital migration that lasts more than 700 Myr.

A comparison between the migration of planets with $M_\mathrm{pl}$ = 0.7 and 4 $ M_\mathrm{J}$ is particularly insightful. Despite the more massive planet depositing more angular momentum into stellar rotation, its orbital decay completes earlier, leaving more time for magnetic braking to erase the tidally enhanced rotation. Eventually, the rotation periods in both simulations converge, and at TOI-2458's age, they reach values consistent with observations. The other two simulations shown in red and blue also allow the star to match the rotation period measurements. However, this time, the stellar rotation period equals its observed value at ages corresponding to either the lower limit from \cite{Subjak} or an upper limit based on the TAMS age (which exceeds the upper age limit from \cite{Subjak}; given that other parameters such as $T_\mathrm{eff}$ and log g suggest that TOI-2458 is an MS star, here we use TAMS as the upper age limit).

\begin{figure}
	\includegraphics[width=\columnwidth]{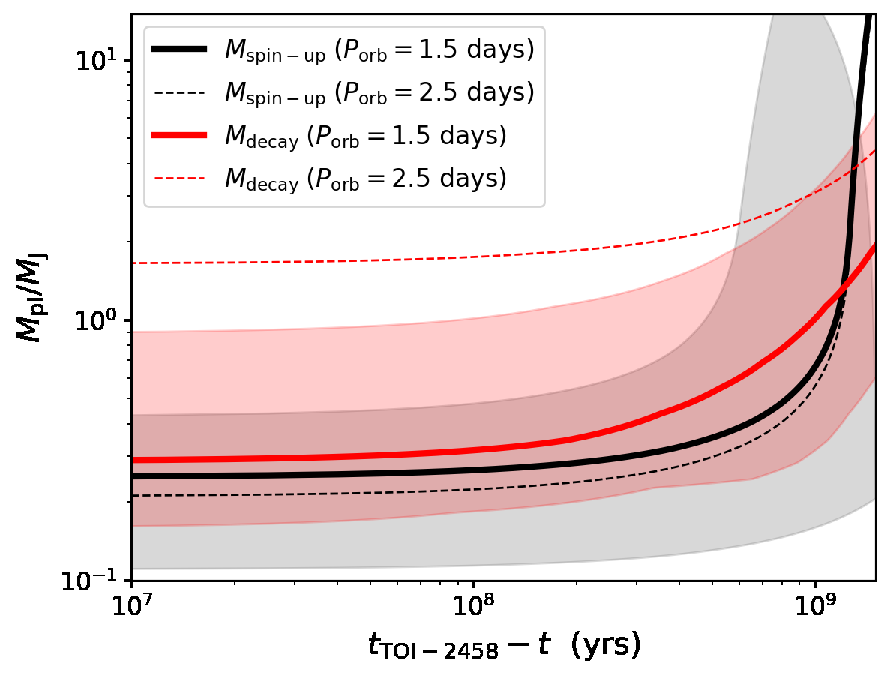}
    \caption{Mass of the engulfed planet as a function of ``lookback'' time. Black lines represent the mass of the planet required to spin-up the host star to the observed present-day rotation ($M_\mathrm{spin-up}$). Red lines indicate the mass of the planet that completes its orbital decay as a result of IGW breaking ($M_\mathrm{decay}$). Solid (dashed) lines correspond to the initial orbital period of 1.5 (2.5) days. Grey and red shaded regions correspond to the uncertainties in $M_\mathrm{spin-up}$ and $M_\mathrm{decay}$ for a 1.5-day planet due to the uncertainties in TOI-2458's measured rotation rate and age, respectively.}
    \label{fig4}
\end{figure}

Having verified that TOI-2458's present-day rotation and age can be explained by a model in which a hypothetical HJ underwent planetary orbital migration driven by IGW breaking, we now explore the parameter space of the problem further. Our four examples, illustrated in Fig.~\ref{fig3}, demonstrate that a planet with higher initial orbital angular momentum must undergo coalescence earlier to allow its host star to spin down to the observed angular velocity at its inferred age. Building on this concept, we compute the mass of the engulfed planet $M_\mathrm{spin-up}$, required to reproduce TOI-2458's observations (hereafter, spin-up mass), as a function of the ``look-back'' time (the time before the present at which the orbital decay finished with the engulfment event) and initial orbital period. In Fig.~\ref{fig4}, the spin-up mass is represented by solid and dashed black lines for fixed initial orbital periods of 1.5 and 2.5 days, respectively. A planet undergoing orbital decay from the shorter periods retains most of its initial orbital momentum until Roche-lobe overflow occurs, which explains the close proximity of these two lines (i.e., the solid and black dashed lines).

Moving to earlier ages (longer look-back times), the spin-up mass monotonically increases. As shown in Fig. \ref{fig2}, the critical mass required for wave breaking also increases with time before the present. The close relation of these trends makes IGW dissipation a promising mechanism for explaining TOI-2458's observed properties. To study this case in more detail, we calculate $M_\mathrm{decay}$, the mass of a planet that completes orbital decay (to engulfment) as a function of age. As we discussed previously, for an initial orbital period of 1.5 days, the orbital decay occurs very rapidly and the planet is engulfed shortly after the onset of IGW breaking. Therefore, in this case, $M_\mathrm{decay} \approx M_\mathrm{crit}$, where $M_\mathrm{crit}$ is the critical planetary mass required for wave breaking, derived by solving $A_\mathrm{nl} = 1$ with $A_\mathrm{nl}$ obtained from equation~(\ref{eq:tide_gw1}). In Fig.~\ref{fig4}, this quantity is shown as a solid red line. From the direct comparison between the solid black and red lines, we find that $M_\mathrm{decay}$ and $M_\mathrm{spin-up}$ almost overlap within the range [0.3, 1] $M_\mathrm{J}$. In contrast, for longer orbital periods, $M_\mathrm{decay}$ and $M_\mathrm{crit}$ do not coincide, as the planet requires a longer time to reach the Roche limit. Consequently, the mass of the decaying planet that started its orbital migration from an orbital period of 2.5 days, shown as the dashed red line, equals the spin-up mass at $M_\mathrm{pl} \sim 4 \; M_\mathrm{J}$. In summary, IGW dissipation is capable of explaining the enhanced spin of TOI-2458 through the previous engulfment of a hypothetical HJ with a mass above 0.3 $M_\mathrm{J}$ and an initial orbital period below 2.5 days. 

Given the current uncertainties in $M_\mathrm{spin-up}$ and $M_\mathrm{decay}$, associated with the precision of rotation rate measurements and age estimates, which are represented by the grey and red shading around the solid lines in Fig.~\ref{fig4}, we cannot derive more definitive constraints on the engulfed HJ. Improved age and rotation rate measurements would be required for us to constrain the properties of this planet further under this scenario. Nevertheless, TOI-2458 and its possible past dynamical evolution constitute an excellent system to constrain IGW dissipation and its implications for spin-orbit evolution.

The dependence of the critical mass on the system's age is a key advantage of the wave breaking scenario. It eliminates the need for precise fine-tuning of the initial orbital period and planetary mass to reproduce the current rotation rate of TOI-2458. Unlike weakly nonlinear processes \citep[e.g.][]{Weinberg} or radiative damping leading to critical layer formation \citep[e.g.][]{Guo23}, which would both cause planetary engulfment to occur too early unless the innermost planet's initial orbital period exceeds the stability threshold for TOI-2458 b, wave breaking initiates sufficiently late so that the tidally enhanced spin-up is observable. This allows us to reproduce the rotation period of TOI-2458 with a previously engulfed planet spanning the entire HJ mass range.

\subsection{GJ 504}

Another intriguing example of a star with enhanced rotation -- compared with expectations for solitary stars -- in the late stages of its MS lifetime is GJ 504. This is a $ \sim 1.22 \; M_\mathrm{\odot}$ star that hosts a wide companion, which is potentially a brown dwarf or a cold Jupiter \citep{D'Orazi2017}. Its reported rotation period of $3.4 \pm 0.25$ days \citep{DiMauro2022} appears remarkably short for its age of $\sim 2.5^{+1}_{-0.7}$ Gyr \citep{D'Orazi2017}, suggesting that a recent planetary engulfment event may have spun up the host star in this system also. This idea was previously examined by \cite{D'Orazi2017} using a constant tidal quality factor model. More recently, \cite{Pezzotti2025} advanced this work by implementing a more realistic tidal prescription, which accounts for the dissipation of equilibrium tides and inertial waves. In this subsection, we focus on the conversion of IGWs to magnetic waves as a possible mechanism enabling rapid planetary migration instead.

\begin{figure}
	\includegraphics[width=\columnwidth]{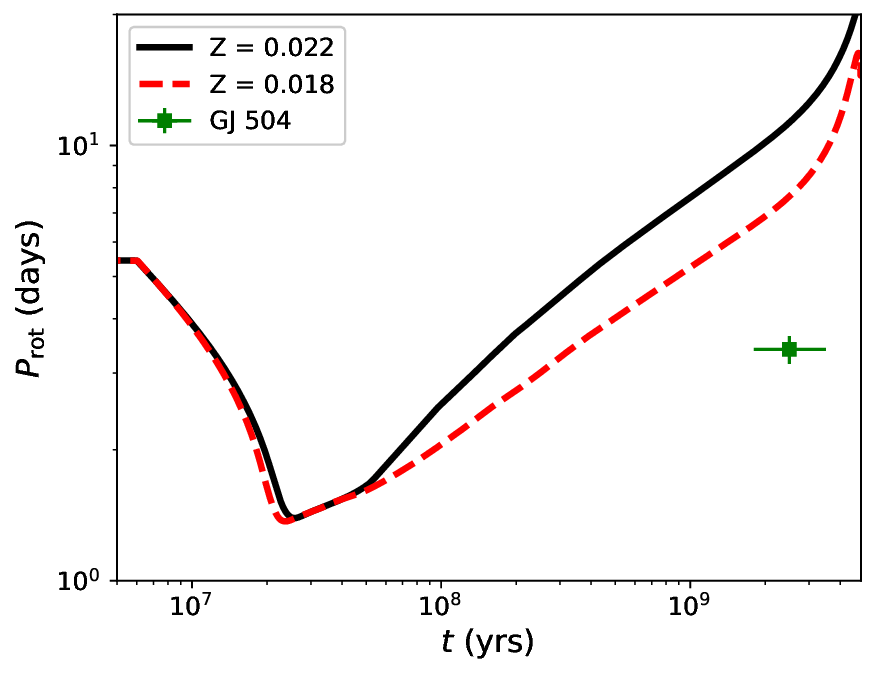}
    \caption{Evolution of stellar rotation period as a function of age for a 1.22 $M_\mathrm{\odot}$ solitary star. Solid black line corresponds to the metallicity from~\protect\cite{D'Orazi2017}; dashed red line represents metallicity from~\protect\cite{Baburaj2025}.
    Green square depicts GJ 504's observed rotation rate (from~\protect\citealt{DiMauro2022}) and age (from~\protect\citealt{D'Orazi2017}).}
    \label{fig5}
\end{figure}

Before examining this case in detail, it is important to stress that stellar rotational evolution near the Kraft break is strongly metallicity dependent, as reported by \cite{AmardMatt}. This dependence arises because the magnetic braking rate, which controls stellar rotation, is determined by the thickness of the convective envelope. In stars as massive as GJ 504, the latter is highly sensitive to the star's chemical composition. According to \cite{D'Orazi2017}, GJ 504 has a supersolar metallicity, with [Fe/H] = 0.22. However, more recent research by \cite{Baburaj2025} suggests a lower value, [Fe/H] = 0.015 and [M/H] = 0.012. Although seemingly small, this difference significantly affects our models' rotational evolution, as shown in Fig.~\ref{fig5}, where a metal-poor model rotates about 1.4 times faster at GJ 504's current age. Therefore, our results for the mass and initial orbital period of the engulfed planet, discussed below, will depend on the adopted stellar metallicity.

\begin{figure}
	\includegraphics[width=\columnwidth]{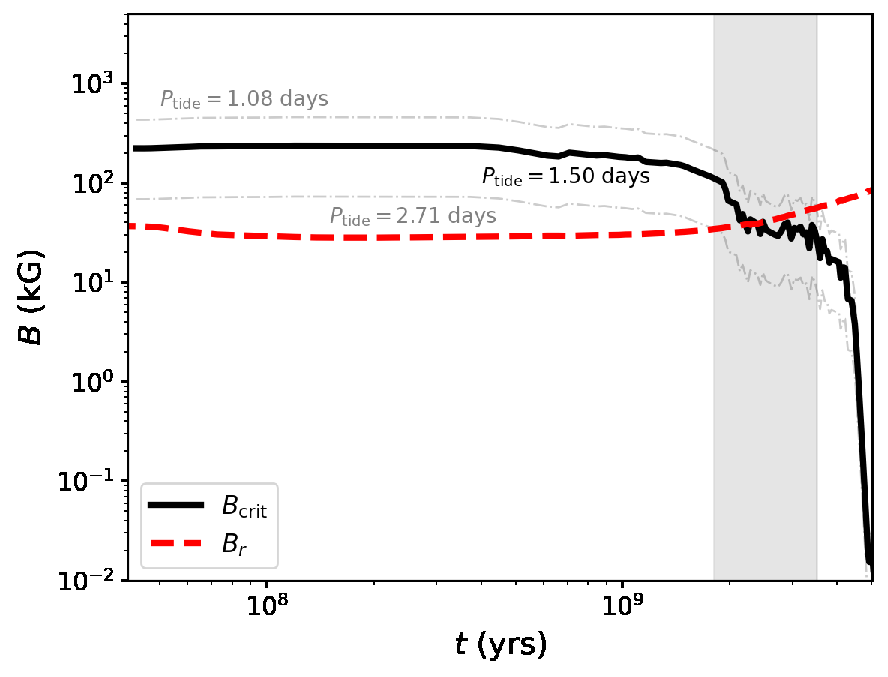}
    \caption{Evolution of the radial magnetic field strength of the convective core (dashed red line) and the critical magnetic field strength (dark solid and dash-dotted grey lines) for GJ 504. Shaded grey region represents the age estimate from~\protect\cite{D'Orazi2017}. Tidal period values used to calculate the critical field strength are given beside the corresponding line.}
    \label{fig6}
\end{figure}

Fig. \ref{fig6} compares the radial magnetic field strength, $B_r$, with the critical magnetic field strength, $B_\mathrm{crit}$, as a function of age for GJ 504 with a metallicity Z = 0.022, which corresponds to [Fe/H] = 0.22. The black solid and grey dash-dotted lines represent tidal periods for which $B_\mathrm{crit}$ equals $B_r$ at the current inferred age of GJ 504, and at the age limits obtained by \citet{D'Orazi2017}, respectively. The age limits are shown by the grey shaded region. For the lowest stellar age compatible with observations, only planets with $P_\mathrm{tide}
 >2.71$ days can excite IGWs capable of being efficiently dissipated by magnetic wave conversion (i.e., when $B_r$ exceeds $B_\mathrm{crit}$). In contrast, for the oldest possible age, the critical tidal period is 1.08 days. For a model with Z = 0.018, the aforementioned critical tidal periods are instead 2.37 and 0.78 days, corresponding to the lowest and highest age inferred for GJ 504, respectively. Therefore, in the metal-poor model, inward migration is permitted in more compact systems. At the same time, as shown in \cite{L21}, the tidal quality factor and the metallicity are anti-correlated, resulting in a faster migration rate for the metal-rich model.

 \begin{figure*}
\begin{multicols}{2}
    \includegraphics[width=\linewidth]{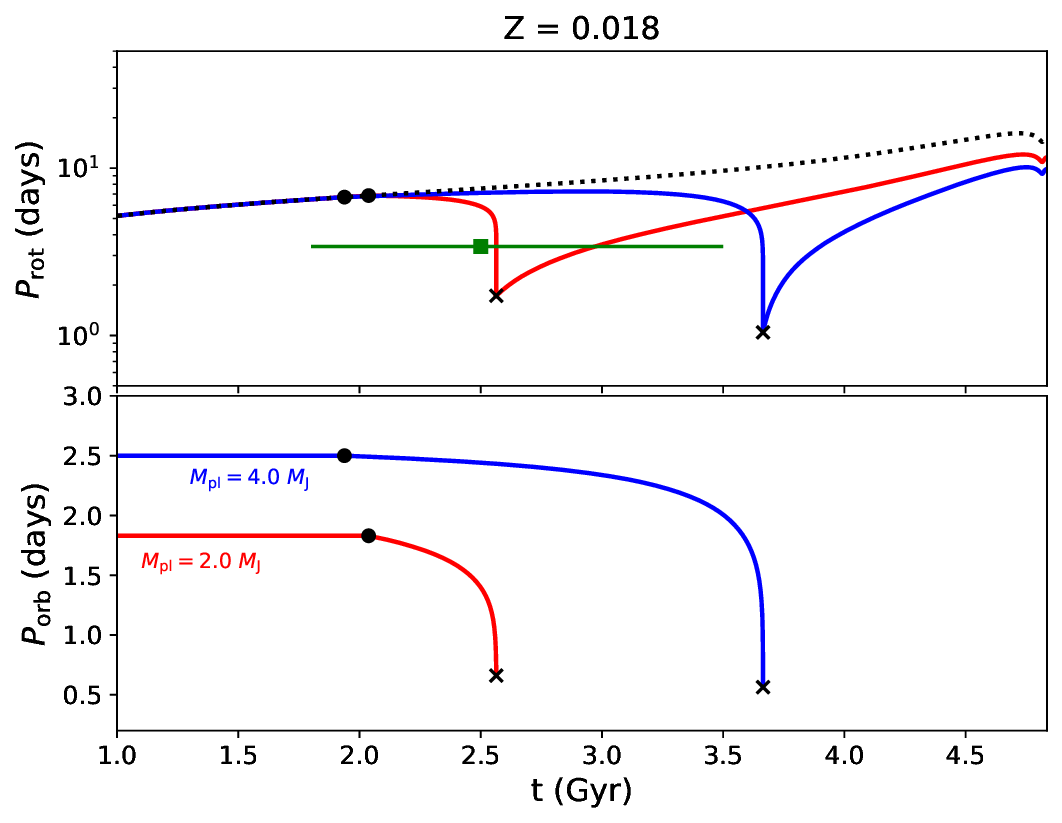}\par 
    \includegraphics[width=\linewidth]{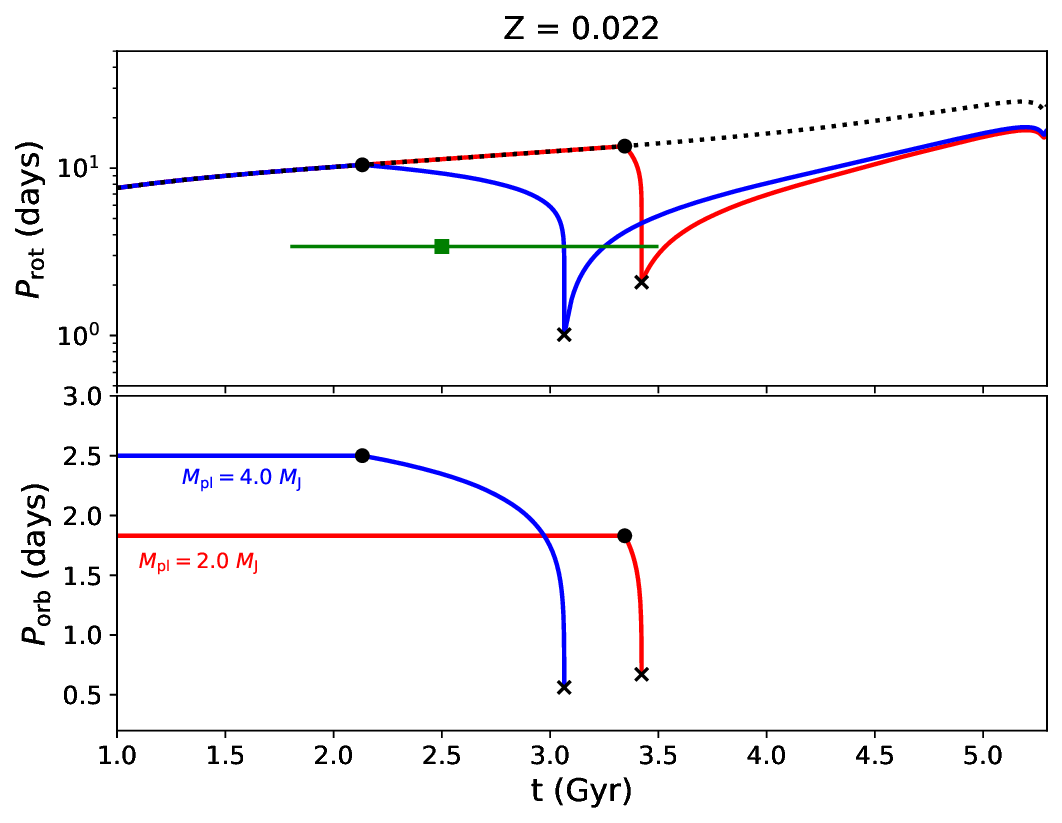}\par 
    \end{multicols}
\caption{Same as Fig. \ref{fig3}, but for GJ 504's model at two metallicities: lower metallicity (left panel) and higher metallicity (right panel).}
\label{fig7}
\end{figure*}

 These trends are illustrated in Fig.~\ref{fig7}, where we compare the orbital evolution of hot Jupiters around GJ 504 for Z = 0.018 (``metal-poor'', left panel) and Z = 0.022 (``metal-rich'', right panel). This shows both the stellar rotation period (top) and the planetary orbital period (bottom) for both models as a function of age for two different planetary masses. For a 2 $M_\mathrm{J}$ planet with an initial orbital period of 1.8 days (red lines), the onset of IGW dissipation is significantly delayed in the metal-rich model. This results in planetary engulfment occurring just too late to account for GJ 504's observed rotation -- and the absence of the planet -- at its present age.
 Conversely, for the metal-poor model, planetary migration begins early enough to align with the observations, despite its slower rate. The impact of the decay timescale becomes crucial for a 4 $M_\mathrm{J}$ planet with an initial orbital period of 2.5 days (blue lines).  As shown in the left panels, the HJ is unable to merge with the metal-poor star before the present day because of the relatively lower dissipation rates in this model. In contrast, the coalescence with the metal-rich star occurs early enough to be consistent with observations. These differences suggest that the region of the initial star-planet separation, for which it is possible to reproduce the age and rotation of GJ 504, varies greatly with metallicity, shifting farther away from the star when moving from a metal-poor to a metal-rich stellar model.

\begin{figure}
\includegraphics[width=\columnwidth]{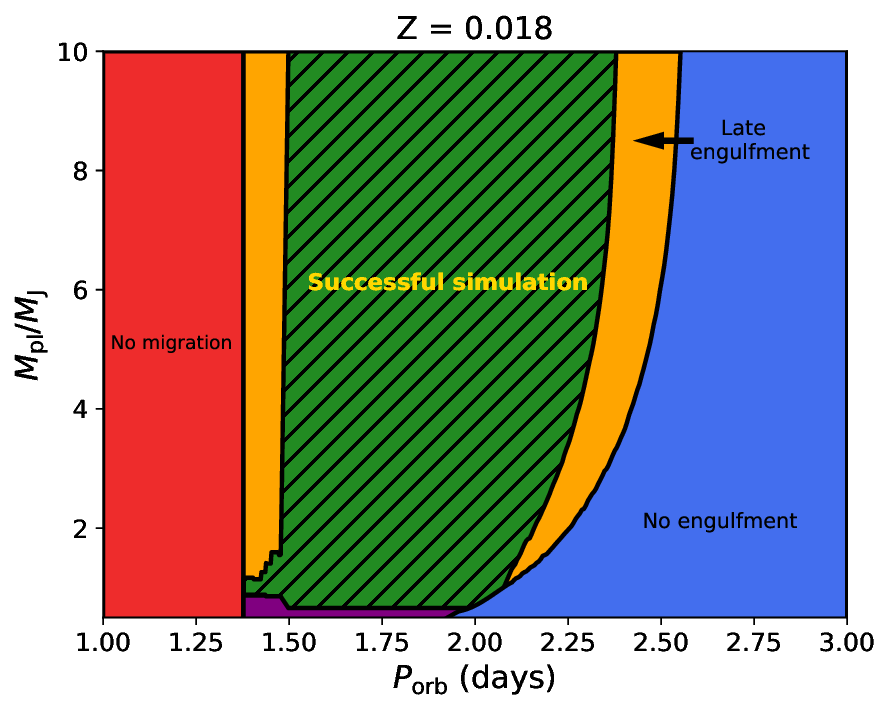}
\includegraphics[width=\columnwidth]{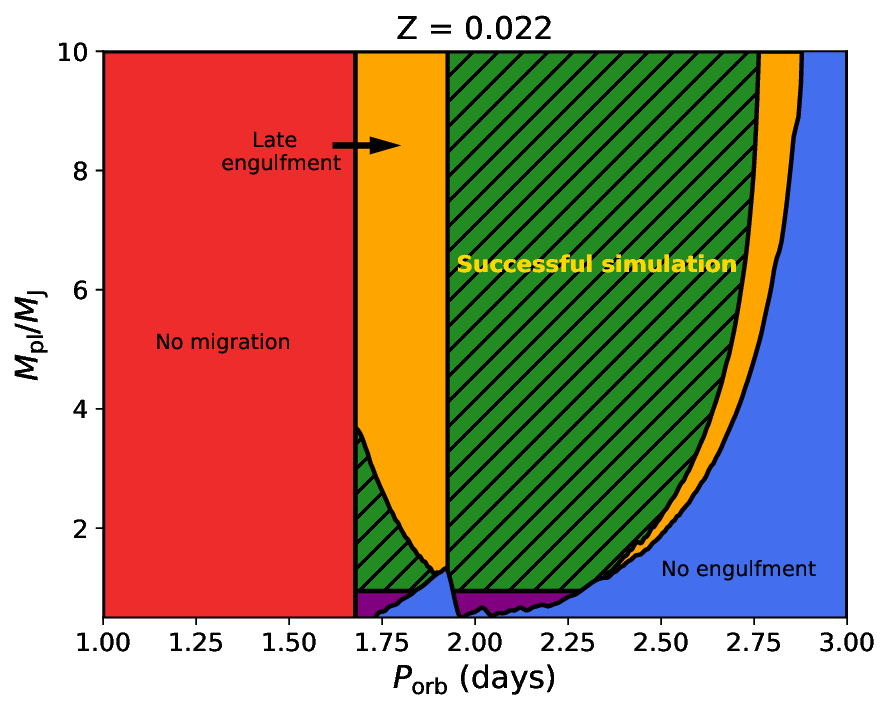}
    \caption{Mass-orbital period diagram for GJ 504. The colors indicate the outcomes of the orbital evolution simulations. The top panel represents the lower-metallicity stellar model and the bottom panel represents the higher-metallicity one. The green hatched region displays the parameters that successfully reproduce GJ 504's current rotation and age. The red region identifies the parameter space where IGW conversion does not occur before a star reaches 3.5 Gyr, the maximum observed age of GJ 504. Planets initially located in the blue region survive until 3.5 Gyr. Planets in the yellow region are engulfed too late for the host star to spin down to GJ 504's observed rotation rate. The mass of planets in the purple region is insufficient to spin up the host star to the required rotation rate.}
    \label{fig8}
\end{figure}

To examine the case of GJ 504 in greater detail, we performed a series of numerical simulations across a grid of initial orbital periods and planetary masses. The grid consists of 300$\times$300 points, uniformly distributed within the ranges $P_\mathrm{orb} = $ 1 to 3 days and $M_\mathrm{pl} = $ 0.5 to 10 $M_\mathrm{J}$. The results of the simulations are presented in Fig.~\ref{fig8}. As before, our simulations begin at the onset of IGW dissipation, ignoring any migration that occurred prior to this point. The parameters that allow us to reproduce GJ 504's present-day rotation following engulfment are represented by the hatched green area.  This region of compatibility varies from $P_\mathrm{orb} \sim 1.5 -2.25$ days, $M_\mathrm{pl} \geq  0.66 \, M_\mathrm{J}$ for the metal-poor model to $P_\mathrm{orb} \sim 1.65 -2.75$ days, $M_\mathrm{pl} \geq  0.95 \, M_\mathrm{J}$ for the metal-rich model of GJ 504. This indicates the significant impact of stellar metallicity on the tidally driven migration. The yellow region, located on either side of the compatibility region, represents scenarios where the planet is also engulfed before 3.5 Gyr, the maximum possible age of GJ 504. However, for the corresponding planets, the engulfment happens too late for the star to spin down sufficiently to match the observed rotation rate. The region shown in purple represents planets with masses that are too low to spin up their host star to GJ 504's observed rotation rate. As demonstrated in Fig.~\ref{fig5}, solitary stars with higher metallicity tend to rotate more slowly than their metal-poor counterparts. This explains why, for Z = 0.022, the engulfment of a higher-mass planet is required to align the stellar spin with observations. The blue region represents planets that merge with their host stars after 3.5 Gyr. Consequently, with parameters in this region, a close-in planet would still be observed around GJ 504 at its current age, contrary to observations. As mentioned previously, the orbital decay timescale is longer for a lower-metallicity model, which explains why the blue region in the top panel is more extended than it is in the bottom panel. Finally, for planets initially located in the red region, the onset of wave conversion occurs after 3.5 Gyr, meaning that no significant migration occurs in these systems before the present day. This region is larger for the higher-metallicity model, which is attributed to the dependence of the critical magnetic field strength, $B_\mathrm{crit}$, on metallicity.

In summary, the results obtained for both lower- and higher-metallicity stellar models indicate that IGW conversion can successfully account for the enhanced rotation observed for GJ 504 across a wide range of planetary masses and initial separations. Compared to the findings of \cite{Pezzotti2025}, who used a migration model based on equilibrium tides and inertial waves, our region of compatibility extends to higher orbital periods and lower planetary masses. However, it should be noted that \cite{Pezzotti2025} assumed slightly different parameters for GJ 504. 

TOI-2458 and GJ 504 represent two possible examples where efficient IGW dissipation leading to the HJ engulfment can explain the faster rotation rates of their host stars over expectations for single stars. They thus provide possible evidence for both the wave breaking scenario in a solar-type star (for TOI-2458) and for the magnetic wave conversion scenario in a hotter star with a convective core (GJ 504). Our hypothesis can be reinforced by exploring the presence of additional observational imprints of planetary engulfment. These signposts of planetary ingestion include, but are not limited to, metal enrichment of the stellar photosphere \citep[e.g.,][although see \citealt{Soliman, Sun} for alternative explanations]{Behmard, Liu, Soares, Lane}, stellar spin misalignment relative to remaining planets \citep{Tokuno}, enhanced magnetic activity \citep{Gehan}, and increased luminosity \citep{Montesinos}.

It is important to note that the initial orbital periods discussed in this section are intended to be determined by the relevant formation scenario as well as by any possible inertial wave dissipation during the first 100 million years of stellar evolution. This early phase of the orbital evolution is strongly dependent on the star's initial spin rate. As demonstrated by \cite{L21}, planets orbiting slowly rotating stars exhibit only minor orbital migration until the onset of IGW dissipation. Conversely, in systems with rapidly rotating stars, inertial wave dissipation plays a substantial role, driving planets inside the corotation radius to much shorter orbital periods. Magnetic braking eventually erases any memory of the initial stellar rotation, preventing us from obtaining an unambiguous picture of the orbital evolution during the earliest stages of the system's lifetime. Nonetheless, we expect inertial waves to significantly influence the orbital evolution of the hypothetical planets discussed in this section, especially around GJ 504. Their action will allow planets with sufficiently long orbital periods to undergo migration prior to the onset of IGW dissipation. Investigating the effects of inertial waves will be a vital aspect of future research.

Finally, the results obtained in this section rely on the robustness of the adopted magnetic braking model from \citet{Matt}. We note that the more recent magnetic wind parametrization by \citet{Spada} couples the braking torque to the moment of inertia of the convective envelope. Their parametrization suggests a weaker spin-down torque for stars in the mass range of TOI-2458. For stars with masses similar to GJ 504, however, both studies yield comparable estimates (see their Fig. 8). A reduction in magnetic wind braking for TOI-2458 would shift the black lines in Fig.~\ref{fig4} to the right, permitting convergence with the present-day rotation for more massive HJs at short orbital periods that finish their tidal decay earlier. At the same time, lower-mass planets would need to be initially located closer to the star, as their late engulfment would not leave the host enough time to spin down to the observed rotation rates (for instance, the 2 $M_\mathrm{J}$ planet shown in Fig.~\ref{fig3} will not be able to reproduce observations). Nevertheless, the uncertainties in the age of TOI-2458 exceed the differences between the wind models, precluding a detailed exploration of how this effect influences the size and shape of the parameter space region consistent with the present-day rotation. 
\section{Impact of internal gravity waves on the evolution of hot Jupiter population}
\label{sec:population}

Evidence for orbital decay has been found for only a handful of HJ systems \citep[e.g.][]{Korth2023,Yeh2024,Adams2024}, and WASP-12 b remains the only planet whose inward migration has been confirmed by multiple studies \citep[e.g.][]{Yee2020,Turner2021,Maciejewski2024,Alvarado2024}, but we can still explore the observational impact of IGW dissipation 
on the HJ population as a whole. In this section, we demonstrate that our models of IGW dissipation (in both solar-type and hotter stars) can explain key statistical patterns in the distribution of HJ systems. In the following, we will formulate the main ideas of our planetary population synthesis approach.

\subsection{Main concept of our approach}
\label{sec:approach}
In \cite{L21,L23}, we showed that HJs undergo two migration phases: an early phase dominated by inertial wave dissipation, which takes place before and shortly after the zero-age main-sequence (ZAMS) when the star is rotating more rapidly, followed by a phase dominated by IGW dissipation throughout the second half of the MS lifetime. Between these two phases, when the star has spun down sufficiently by magnetic braking, inertial waves are typically no longer excited, and wave breaking has not yet occurred so that IGW dissipation is not efficient. Tidal migration is then stalled because equilibrium tide dissipation (the remaining mechanism included in those calculations) is too weak to affect the planetary orbits during the MS \citep[due to the frequency-reduction of the effective viscosity in the fast tides regime, following, e.g.~][]{DBJ2020b,Barker,Nils2023}. Assuming aligned and circular orbits for HJs and neglecting magnetic star-planet interactions, these planets therefore exhibit negligible migration after inertial wave dissipation ceases prior to the onset of IGW dissipation. For every observed HJ system with a known age estimate, we can calculate the time when its orbital dynamics would become unfrozen due to the onset of efficient IGW dissipation according to \S~\ref{subsec:mechanisms}. Based on these results, we construct two sub-samples: non-migrating (young) and migrating (old) HJs, after removing very young HJs that may not have finished an early phase of their migration due to inertial waves. Ideally, we would like these two subsamples to have distinct statistical properties reflecting their dynamical states, thereby allowing IGW dissipation theories to be tested. At the same time, these subsamples are necessarily connected, as young (non-migrating) HJs will eventually become old (migrating) ones. 

Furthermore, for every young HJ, we can predict its orbital evolution after the onset of IGW dissipation and determine its orbital period for any given age using the approach described in \S~\ref{subsec:evolution}. Based on the obtained results, we will attempt to recreate the distribution of old planets based on the distribution of young ones, which are treated as the initial population. This approach will require applying the HJ formation history (HJFH) to reassign systems' ages. Our first HJFH option is based on the star formation history (SFH) in the Galactic thin disc from \cite{Mor2019} calibrated using the Stilism extinction map \citep{Lallement2018}. Although HJFH and SFH are not necessarily equivalent, the model by \cite{Mor2019} was shown to align with the ages of star-planet systems from the Kepler field \citep{Bouma2024}, which justifies our choice. The second option we consider is a uniform HJFH between 0 and 10 Gyr. By utilizing the adopted HJFH and planetary orbital migration simulations, we derive the orbital period of a synthetic system. At the same time, the planetary and stellar masses of our synthetic system are set to the parameters of the observed parent system. If IGW dissipation is the dominant mechanism for planetary migration (after 100 Myrs), then our new population should be a good approximation of the distribution of old systems. This experiment will serve as a way to test our current understanding of planetary orbital migration.

\subsection{Sample selection}
\label{subsec:sample}
Our HJ sample is primarily constructed using stellar parameters from \cite{Swastik24}. Their reported host star ages, masses, and metallicities were derived through an isochrone fitting method utilizing MIST isochrones. To refine our sample, we implemented the following selection criteria:
\begin{enumerate}
    \item Stellar mass $M_{\star}$ = 0.7 -- 1.5 $M_{\odot}$, effective temperature $T_\mathrm{eff}$ = 4500--7000 K, and surface gravity log g > 4.0.
    \item Planetary mass $M_\mathrm{pl}$ = 0.2 -- 10 $M_\mathrm{J}$, and orbital period $P_\mathrm{orb}$ < 5 days.
    \item The lower limit on stellar age is within the range [0.1,10] Gyr.
    \item For stars with $M_{\star} < 1.15 \; M_{\odot}$  ($M_{\star} \geq 1.15 \; M_{\odot}$), the age uncertainty is below 3 (1.5) Gyr.  
\end{enumerate}

A total of 218 systems satisfy the aforementioned criteria. The first condition ensures that all host stars are MS stars. The second condition constrains our sample to include only HJ systems with sufficiently compact orbits to be affected by IGW damping. The third criterion removes systems that are likely to be undergoing an early active phase of orbital migration driven by inertial waves. Furthermore, we exclude systems older than 10 Gyr, as our HJFH model has a maximum age of 10 Gyr. The fourth condition allows us to exclude systems with poorly constrained ages.

Our sample is further extended by 102 additional systems, whose parameters, retrieved from the NASA Exoplanet Archive, also satisfy the above criteria. When multiple data entries are available for a given system, we select the one corresponding to the smallest age uncertainty.  

 \begin{figure*}
\begin{multicols}{2}
    \includegraphics[width=\linewidth]{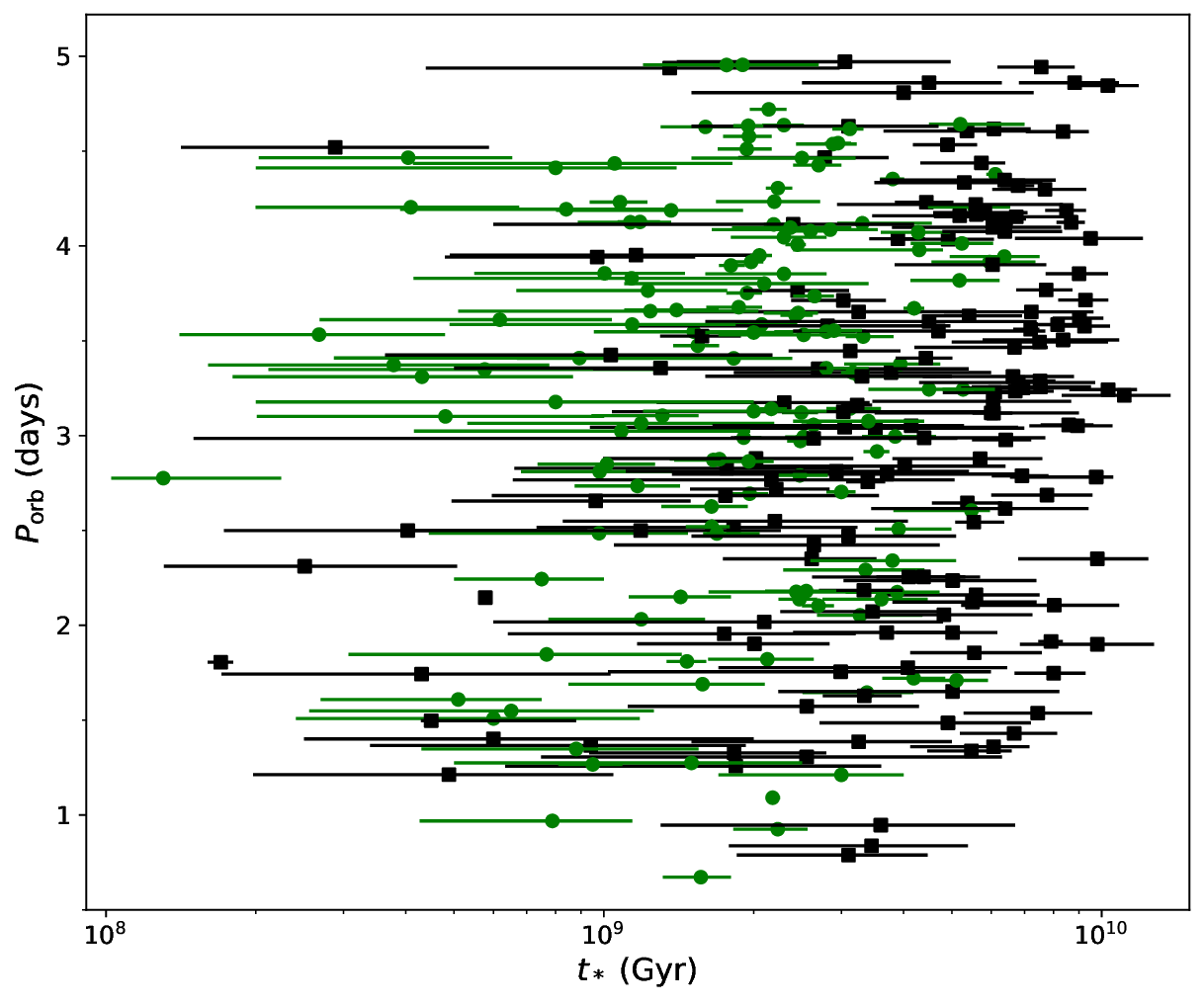}\par 
    \includegraphics[width=\linewidth]{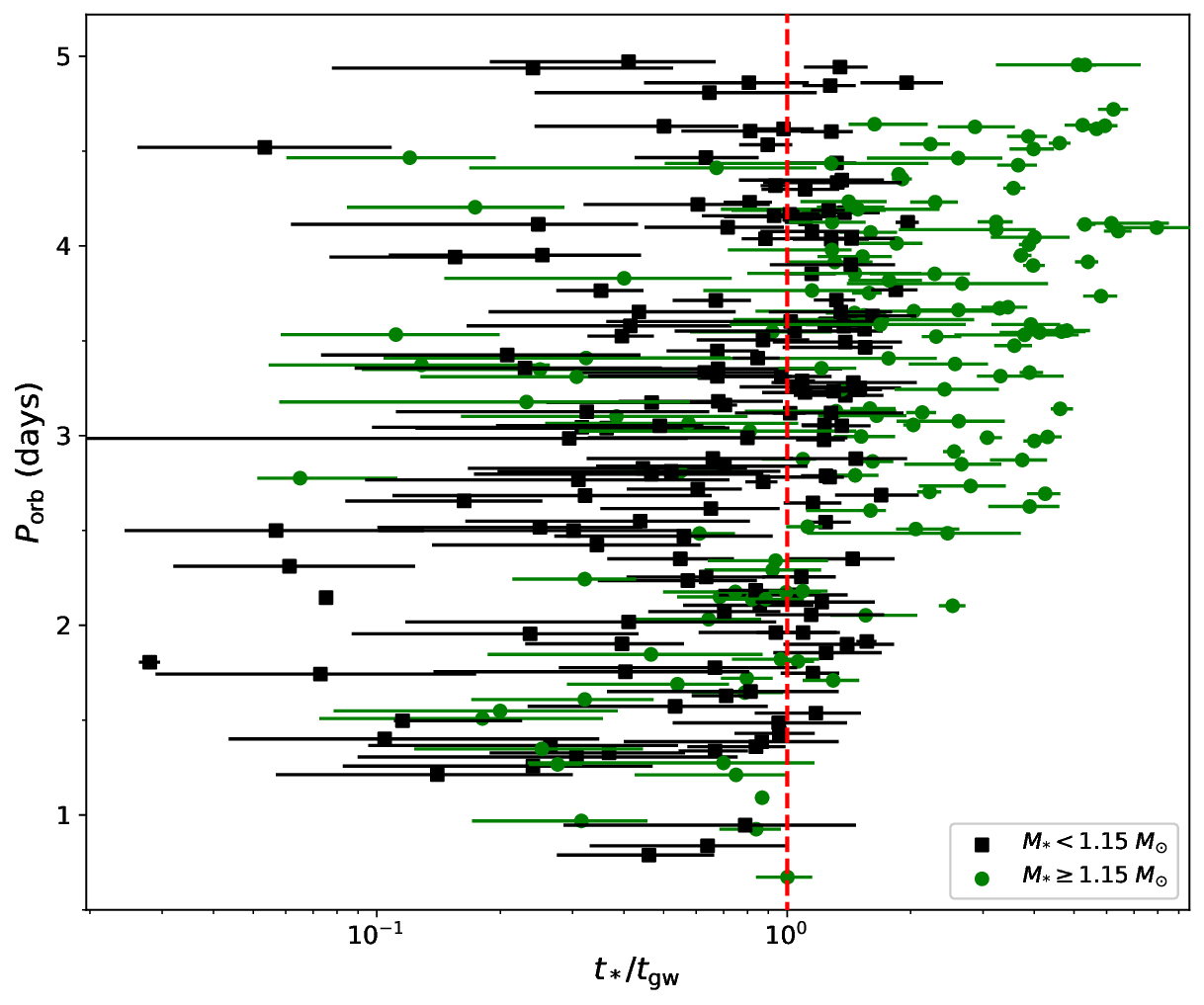}\par 
    \end{multicols}
\caption{Distribution of the observed HJ systems in the age-orbital period diagram. Ages are shown in Gyr in the left panel and normalized to the onset age of IGW dissipation in the right panel. Systems where the wave breaking criterion is applied are shown with black squares, while those where the wave-conversion criterion is adopted are represented by green squares. The red dashed line separates systems with non-migrating planets from those where IGW dissipation is ongoing.}
\label{fig9}
\end{figure*}

\subsection{Young and old observed hot Jupiter systems: overall comparison}
For each host star in our sample, we computed a stellar model following the prescriptions given in Section \ref{subsec:models}. Utilizing the systems' observed parameters and the methods outlined in Section \ref{subsec:mechanisms}, we determined the ages corresponding to the onset of IGW dissipation, $t_\mathrm{gw}$. For stars with $M_{\star} \geq 1.15 \; M_{\odot}$, our models retain a convective core for most of their MS lifetimes, necessitating the application of the magnetic wave conversion criterion because wave breaking is not expected until the emergence of a radiative core later on. Conversely, for lower-mass stars, we assume that IGW dissipation proceeds only through wave breaking when equation~(\ref{eq:tide_gw2}) is satisfied. Furthermore, we neglect stellar rotation when calculating the tidal period $P_\mathrm{tide} \approx P_\mathrm{orb}/2$.
 
We refined our sample by excluding the 24 systems where wave breaking was initiated either before 100 Myr or after 10 Gyr. This leaves us with a total of 296 systems, which are presented in the age-orbital period diagram in Fig.~\ref{fig9}. In the left panel, the ages are given in Gyr, while in the right panel they are normalized to $t_\mathrm{gw}$. This normalization allows us to distinguish between non-migrating and migrating planets, which are located to the left and right of the dashed red line (in the right panel), respectively.

\begin{figure}	\includegraphics[width=\columnwidth]{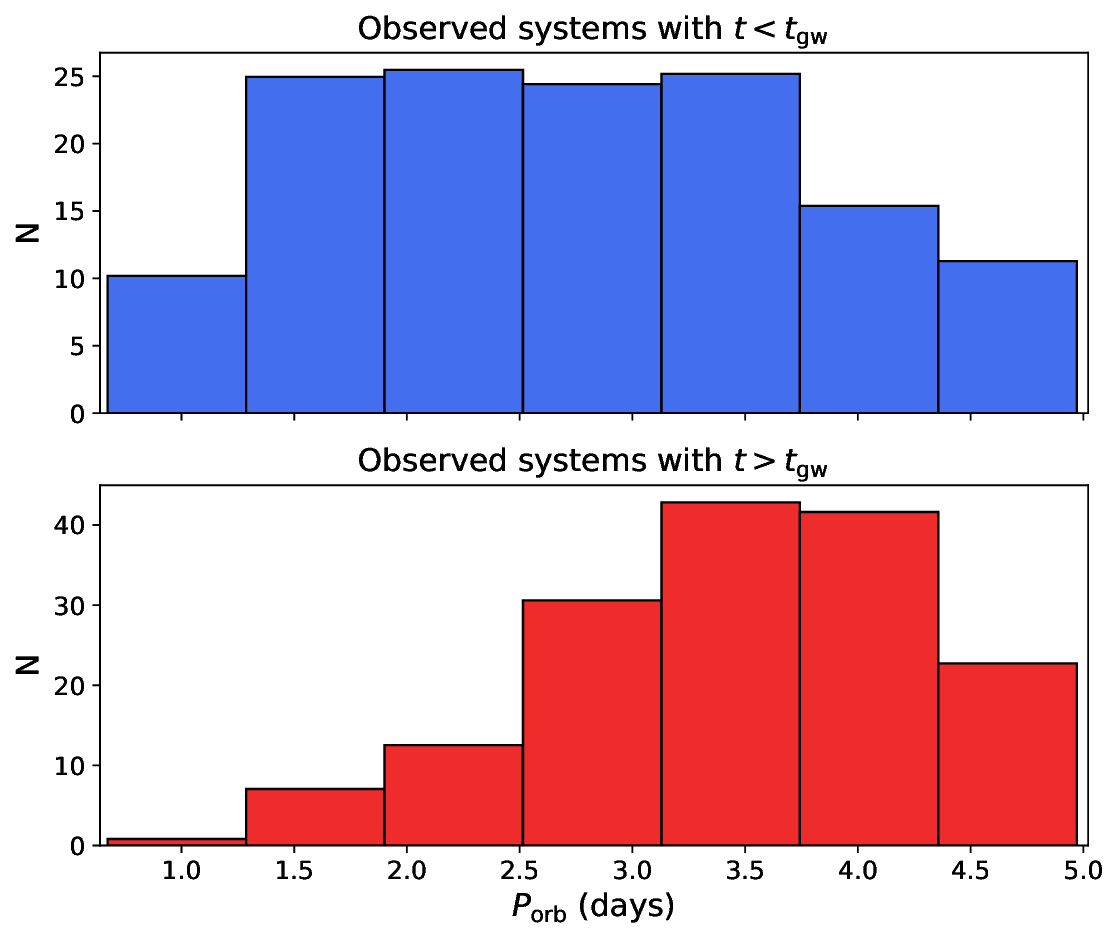}
    \caption{Orbital period distribution of the observed HJ population. Top panel: non-migrating (young) HJs. Bottom panel: migrating (old) HJs}
    \label{fig10}
\end{figure}

A striking contrast is immediately visible between planets with $t / t_\mathrm{gw} < 1$ and those with $t / t_\mathrm{gw} > 1$. Specifically, in Fig.~\ref{fig9}, the orbital periods of the former are evenly distributed, while higher-period HJs are more common among the latter. For systems with a star that has a convective core, this can be explained by the correlation $B_\mathrm{crit} \sim P_\mathrm{tide}^2 \sim P_\mathrm{orb}^2$. However, the same shift toward higher orbital periods is observed for systems with stellar masses below $1.15 \; M_{\odot}$, where $t_\mathrm{gw}$ is only weakly sensitive to the tidal period ($A_{\mathrm{nl}}\propto P_{\mathrm{orb}}^{1/6}$) but is much more strongly correlated with the planetary mass and the system's age. Looking solely at stars with radiative cores in the observed sample, we found that none of the 15 systems with an orbital period of less than 1.5 days permit wave breaking according to equation~(\ref{eq:tide_gw2}). The shortest-period system with $t / t_\mathrm{gw} > 1$ and a stellar mass less than $1.15 \; M_{\odot}$ is WASP-36, which has an orbital period of 1.54 days. These results indicate that IGWs may have had a substantial effect on the HJ population. Of the 296 HJ systems in our sample, 119 (40\%) are predicted to undergo coalescence before the host star reaches the TAMS or 10 Gyr (whichever comes first).

The same trend is observed in Fig.~\ref{fig10}, which compares the orbital period distributions of both HJ subsamples (with either $t / t_\mathrm{gw} < 1$ or $t / t_\mathrm{gw} > 1$). Notably, the number of HJs with $t / t_\mathrm{gw} > 1$ peaks at an orbital period of $3.5 -4$ days. As shown in \cite{L21}, this range separates the inner region of the parameter space, where IGW dissipation can lead to orbital decay, from the outer region, where the systems are unable to be tidally engulfed before TAMS.

\begin{figure}	\includegraphics[width=\columnwidth]{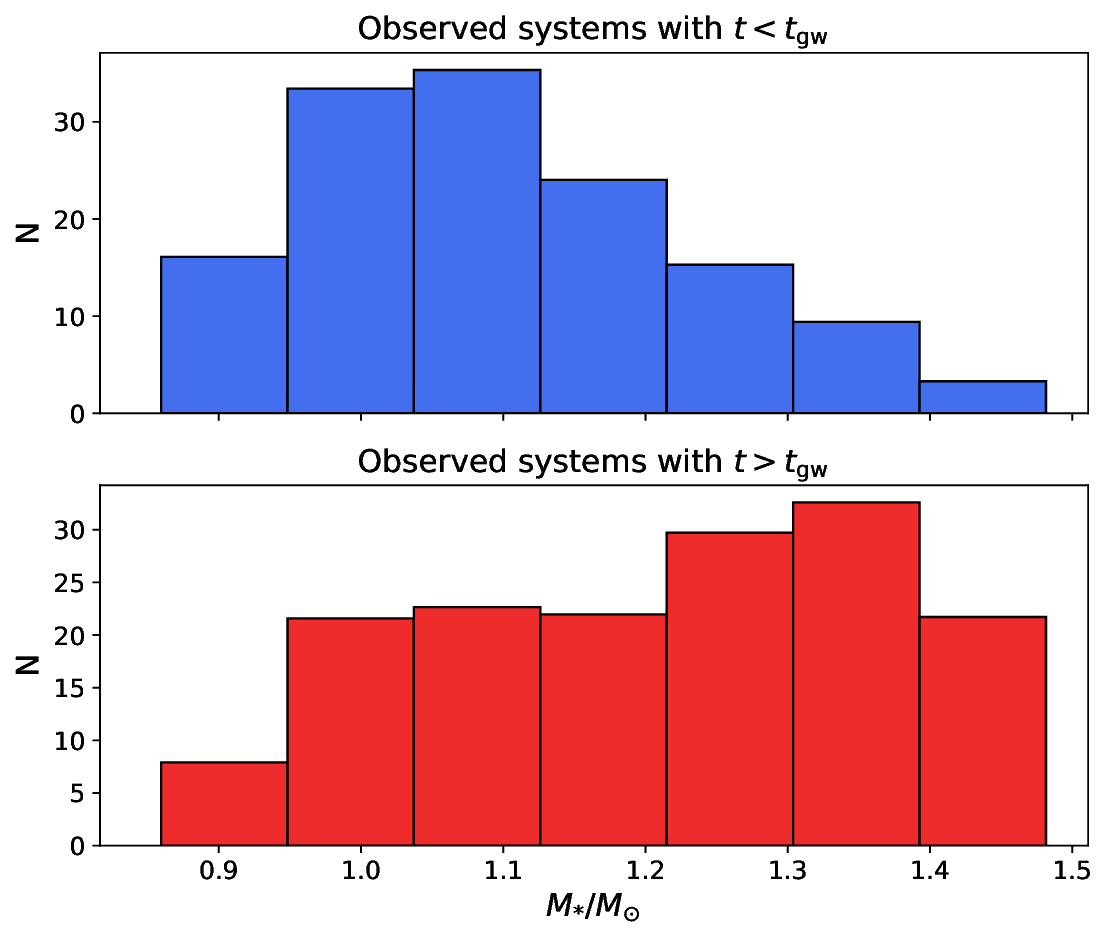}
    \caption{Host star mass distribution of the observed HJ population. }
    \label{fig11}
\end{figure}

In Fig. \ref{fig11}, we compare the stellar mass distributions of young and old HJ hosts. After applying the additional restriction on $t_\mathrm{gw}$ described earlier in this subsection, all systems with a stellar mass below 0.85 $M_{\odot}$ were filtered out. As we discussed previously, IGW dissipation is more likely to be ongoing for systems with a star having a convective core than for those with a radiative core. As a result, systems with $M_{\star} < 1.15 \; M_{\odot}$ dominate the subsample of young systems. Among the old systems, the numbers of stars with a radiative core and those with a convective core are comparable.

\begin{figure}	\includegraphics[width=\columnwidth]{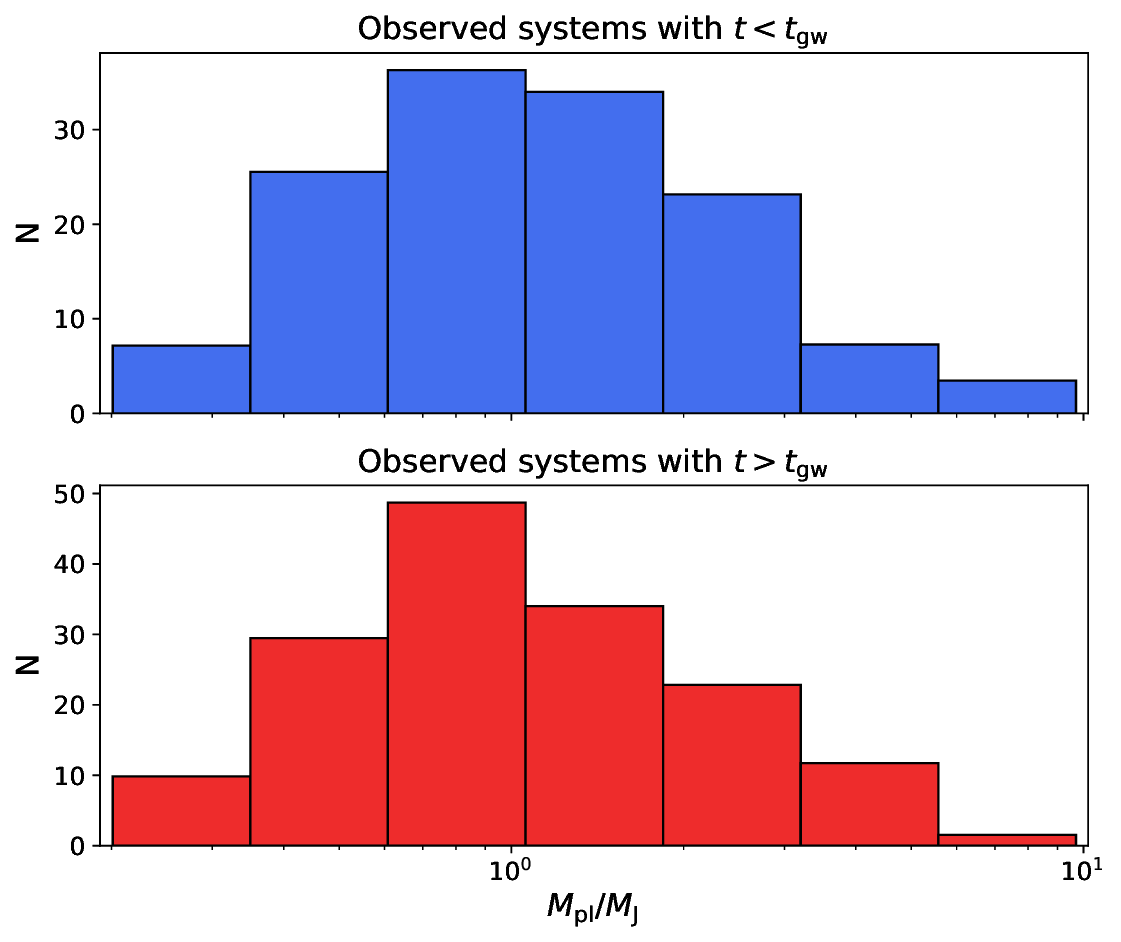}
    \caption{Planetary mass distribution of the observed HJ population. In contrast to Figs. \ref{fig10} and \ref{fig11}, planetary mass is shown on a logarithmic scale.}
    \label{fig12}
\end{figure}

The impact of planetary mass on the age distribution relative to $t_\mathrm{gw}$ is two-fold. First, wave breaking begins earlier for more massive planets, which would lead to a higher proportion of massive HJs in the subsample of planets with $t>t_\mathrm{gw}$. However, massive planets also undergo coalescence earlier due to a faster migration rate once efficient IGW dissipation occurs. This second effect would shift the distribution of old HJs toward lower planetary masses. As shown in Fig.~\ref{fig12}, the decrease in the number of HJs with $M_\mathrm{pl} > 1 \; M_\mathrm{J}$ is slightly steeper for systems with $t<t_\mathrm{gw}$; however, the overall planetary mass distributions are comparable between the two subsamples.

\subsection{Population synthesis model}
In the present work, we aim to recreate the observed decline in the number of HJs with $t > t_\mathrm{gw}$ at very short orbital periods using a population synthesis approach. The resulting synthetic population will comprise systems with orbital periods below 4 days, where the effect of inward migration is pronounced. Our simulations will also utilize the observed systems with orbital periods in the range $4 \leq P_\mathrm{orb} \leq 5$ days. This ``supply region" will serve as a source for planets migrating inwards from larger orbital separations.

Following the approach described in Section \ref{sec:approach}, we generate a synthetic population in three steps. First, we define the initial model, which encompasses all young systems as well as some old systems whose ages, within one $\sigma$ (standard deviation), may be located to the left of the $t = t_\mathrm{gw}$ line. Each of these systems is assigned a weight that determines its contribution to the initial population. This weight is calculated taking into account three factors. 

To convert the observed systems to our initial population, we perform a bias-free correction by introducing the first factor, $C_1 = \frac{a}{R_{\star}}$, which is the inverse transit probability. As we explained earlier, each system in our sample is characterized by its $t_\mathrm{gw}$. A higher $t_\mathrm{gw}$ is equivalent to a higher likelihood of identifying a given system as a system with $t<t_\mathrm{gw}$. Conversely, observing a system with $t < t_\mathrm{gw}$ at low $t_\mathrm{gw}$ implies a higher number of systems with similar parameters in which IGW dissipation is currently enabled. Thus, the second factor is:
\begin{equation}
C_2 = \frac{\int_{\rm 100 \, Myr}^{10 \, \rm Gyr} \mathrm{SFR}(\tau) \; d\tau}{\int_{\rm 100 \, Myr}^{t_\mathrm{gw}} \mathrm{SFR}(\tau) \; d\tau},
\label{eq:c2}
\end{equation}
 where the integration over a lookback time $\tau$ is performed based on the adopted SFH (or, equivalently, HJFH). The third factor, $C_3 = 1/2 + 1/2 \rm erf\left(\frac{t_\mathrm{gw}-t_{\star}}{\sqrt{2}\sigma}\right)$, with $\rm erf$ the Gauss error function, is introduced specifically for systems whose age errorbars cross the $t = t_\mathrm{gw}$ line. This factor determines the probability that the true age of a system is below $t_\mathrm{gw}$. The final weight is calculated as $C_\star = C_1C_2C_3$.

\begin{figure*}
\begin{multicols}{2}
    \includegraphics[width=\linewidth]{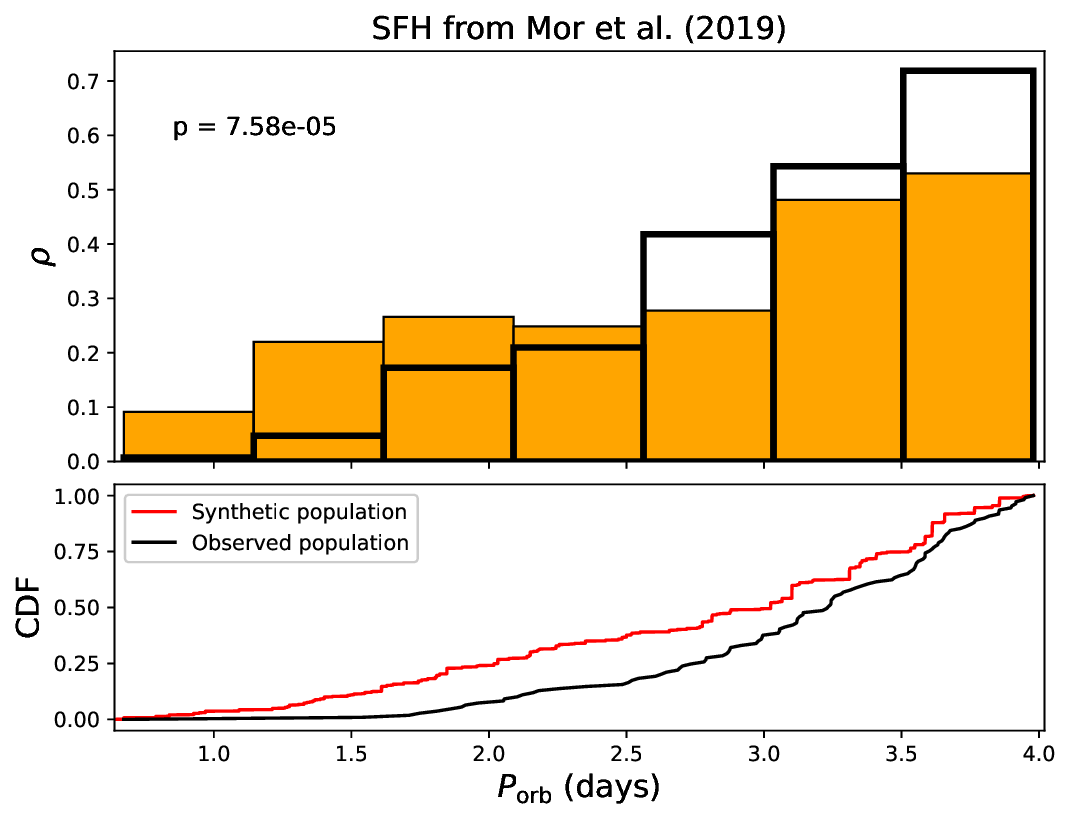}\par 
    \includegraphics[width=\linewidth]{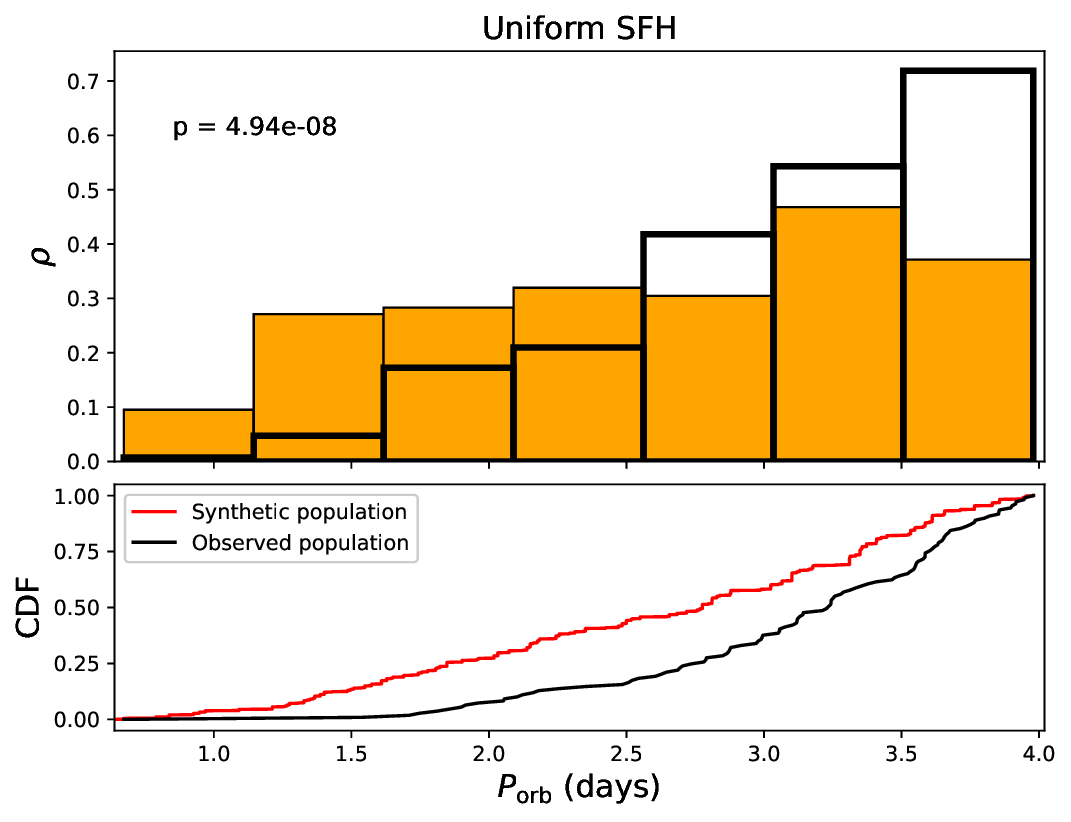}\par 
    \end{multicols}
\caption{Top panel: probability density functions of the orbital period for the synthetic populations (yellow histogram) and for the observed systems with $t > t_\mathrm{gw}$ (black contour line). The synthetic populations are obtained assuming non-evolving orbits. On the left, we print the KS $p$-value quantifying the difference between the two samples. Bottom panel: cumulative distribution functions for the same populations. Left panel: SFH from~\protect\cite{Mor2019}. Right panel: uniform SFH. This figure demonstrates that orbital migration is necessary to reproduce the observed distribution.}
\label{fig13}
\end{figure*}

Second, for each system in our initial population, we sample its age $N = 2\times 10^4C_\star$ times from the adopted SFH, retaining only values less than both the MS lifetime of the corresponding host star and the adopted upper age limit of 10 Gyr. Before applying our migration model, we assume a scenario in which the orbits of our synthetic systems remain fixed and assign each synthetic system belonging to the same parent system its present-day orbital period. In Fig.~\ref{fig13}, we display the probability density functions corresponding to the observed (black contour line) and the synthetic (yellow histogram) population of old HJs based on a no-migration scenario. Without migration, our synthetic population cannot match the observational statistics. The two adopted HJFHs produce slightly different shapes of the orbital period distribution, both appearing too flat compared to observations. To quantify this divergence, we perform two-sample Kolmogorov-Smirnov (KS) tests and display the resulting $p$-values at the top left of each plot. The cumulative distribution functions, shown at the bottom of Fig.~\ref{fig13}, confirm that the no-migration model generates systems with orbital periods that are on average shorter than those in the observed sample.

\begin{figure}	\includegraphics[width=\columnwidth]{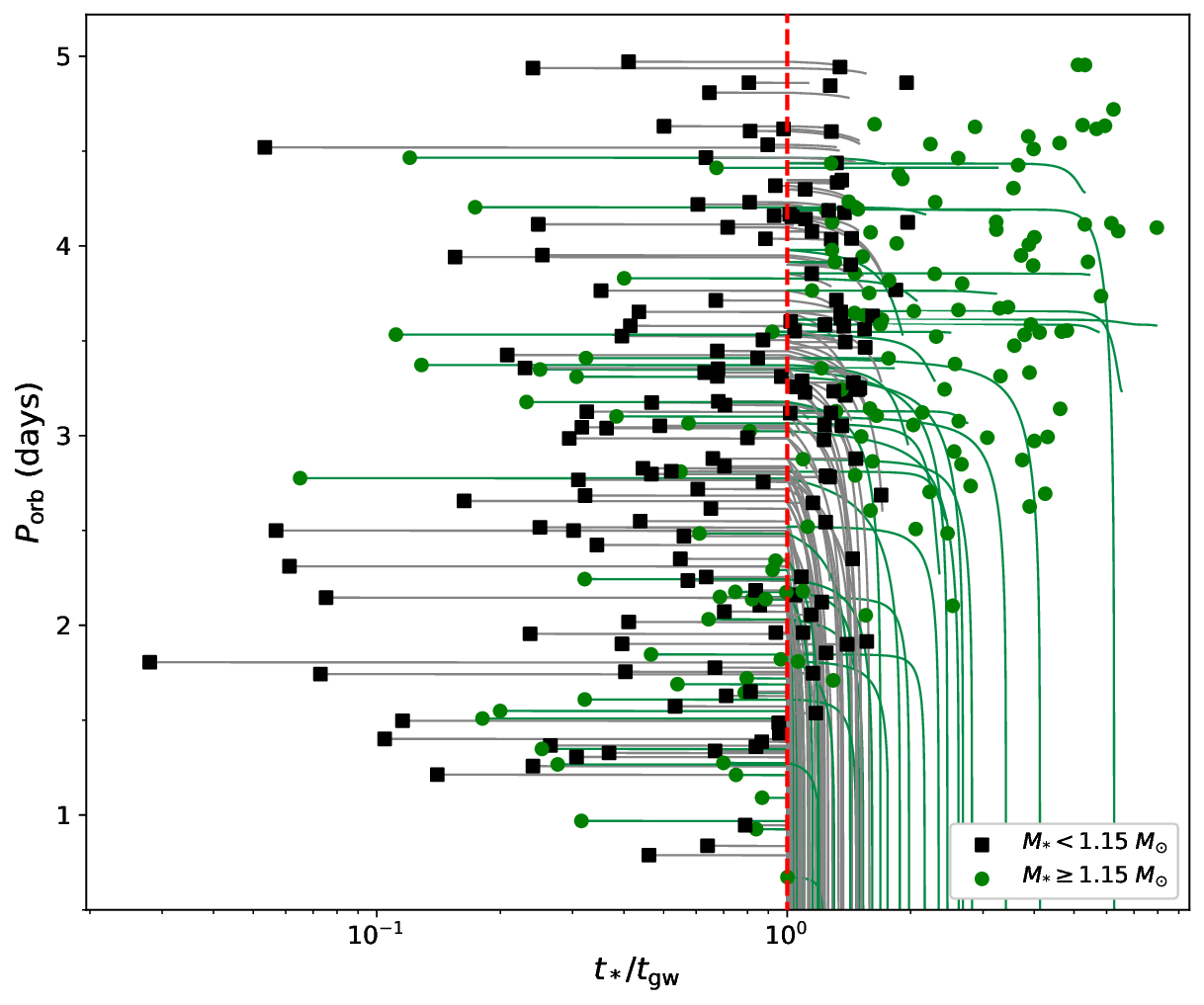}
    \caption{Same as the right panel of Fig. \ref{fig9}, but with the lines illustrating the migration of young Jupiters.}
    \label{fig14}
\end{figure}

Third, we simulate the orbital migration of our initial population and use the obtained results to derive the orbital period at a given age. The trajectories of the observed young planets, orbiting stars with radiative and convective cores, are shown in Fig.~\ref{fig14} with grey and green lines, respectively. After computing a synthetic system's orbital period, we estimate its transit probability. We then perform a transit test and exclude systems that fail, that is, those for which a random number sampled from a uniform distribution between 0 and 1 is above their transit probability. Planets that fill their Roche lobes at the simulated age are counted as engulfed. The remaining systems form our synthetic transiting population.

\subsection{Results}

Planetary orbital migration can lead to the coalescence of star-planet systems with the smallest separations, leaving imprints on the orbital period distribution of the surviving planets. Our primary task is to determine whether IGW dissipation can reproduce the observed population of HJs in old systems where wave breaking or wave conversion are expected.

\begin{figure*}
\begin{multicols}{2}
\includegraphics[width=\columnwidth]{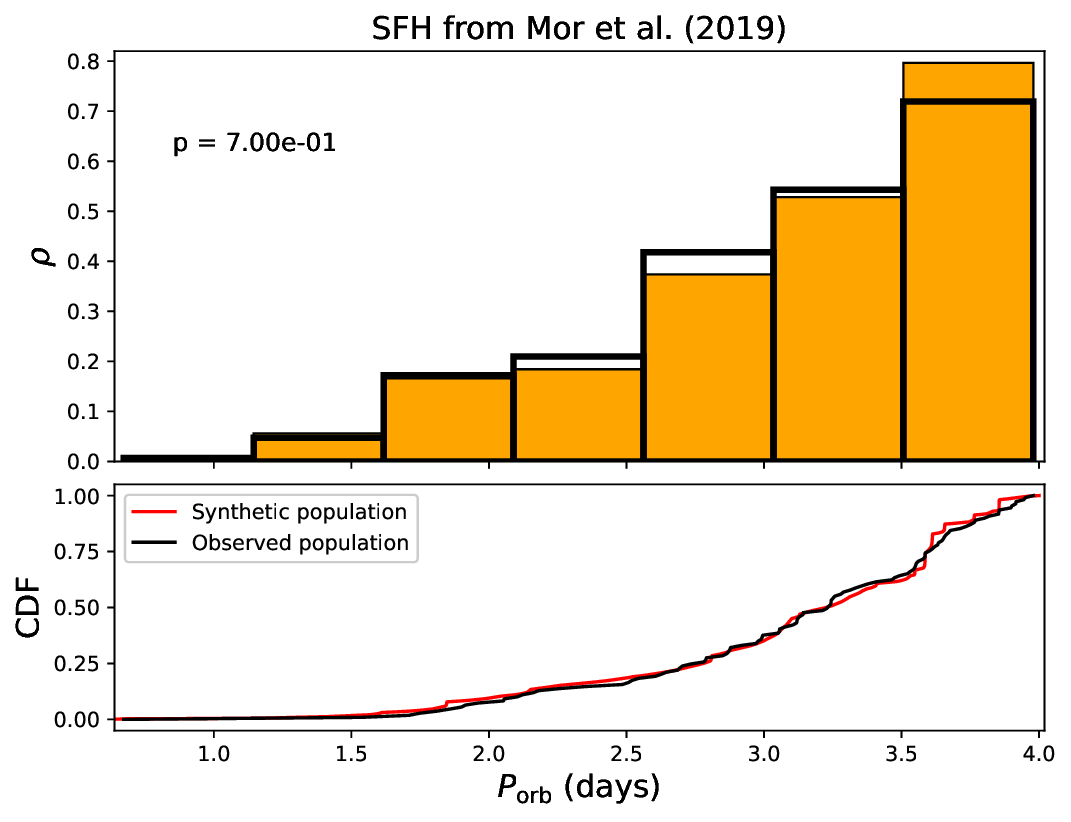}\par 
\includegraphics[width=\columnwidth]{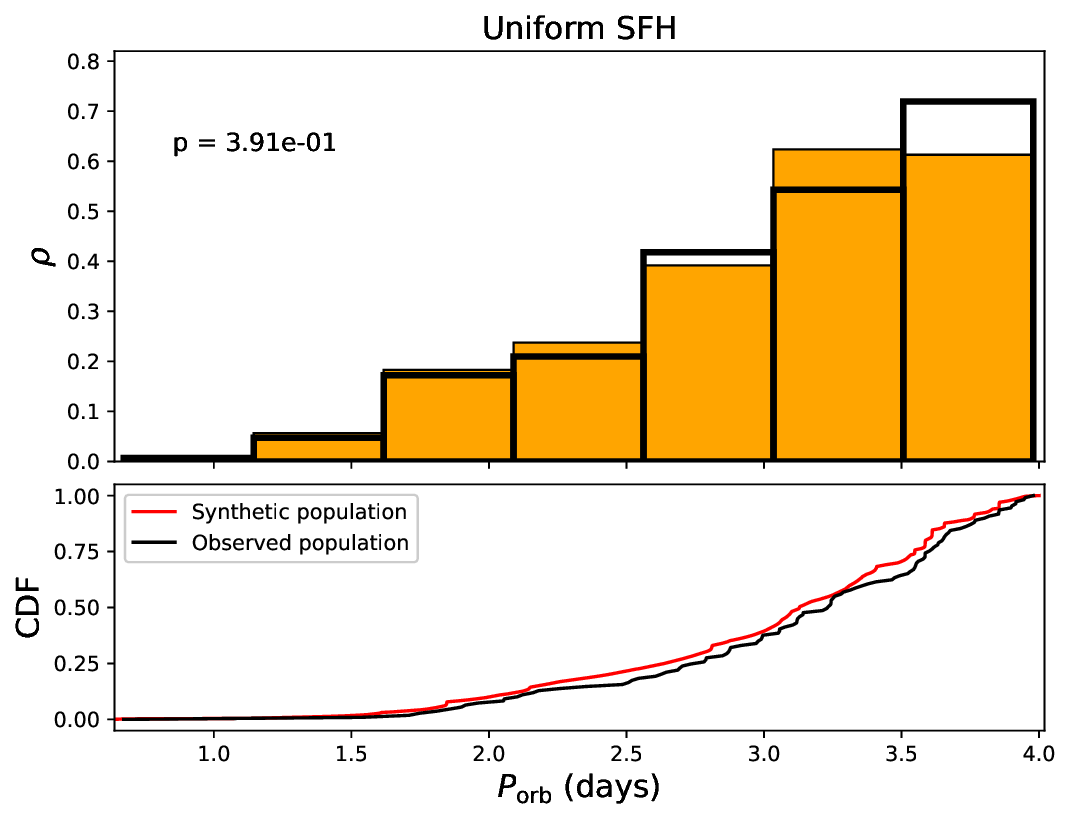}\par 
\end{multicols}
    \caption{The same as Fig. \ref{fig13}, but synthetic populations are generated assuming IGW-driven migration. This shows that IGW-driven migration can reproduce the observed distribution of HJs.}
    \label{fig15}
\end{figure*}

This question is addressed in Fig.~\ref{fig15}, which compares the orbital period distributions of the observed old systems (black contour line) with the synthetic population of old systems created from the observed distribution of young systems as initial conditions (yellow histogram). The agreement between the two populations is very good, as further validated by the results of the KS test, presented in the top left corner, and the comparison of the cumulative density functions, shown at the bottom. Both the SFH from \cite{Mor2019} and a uniform SFH produce synthetic populations whose orbital period distributions are statistically indistinguishable from the observed HJ subsample with $p > 0.05$. Notably, the SFH from \cite{Mor2019} results in convergence at a higher confidence level. This result provides new evidence that IGW dissipation is responsible for sculpting the population of old HJ systems.

Given that planetary migration does not directly affect the distribution of planetary and host star masses, and considering the limited number of HJs in our observed sample, we should not expect the synthetic and real stellar and planetary mass distributions to be identical. Nevertheless, it is informative to assess the degree of similarity between these distributions to highlight areas for further improvement.

\begin{figure*}
\begin{multicols}{2}
\includegraphics[width=\columnwidth]{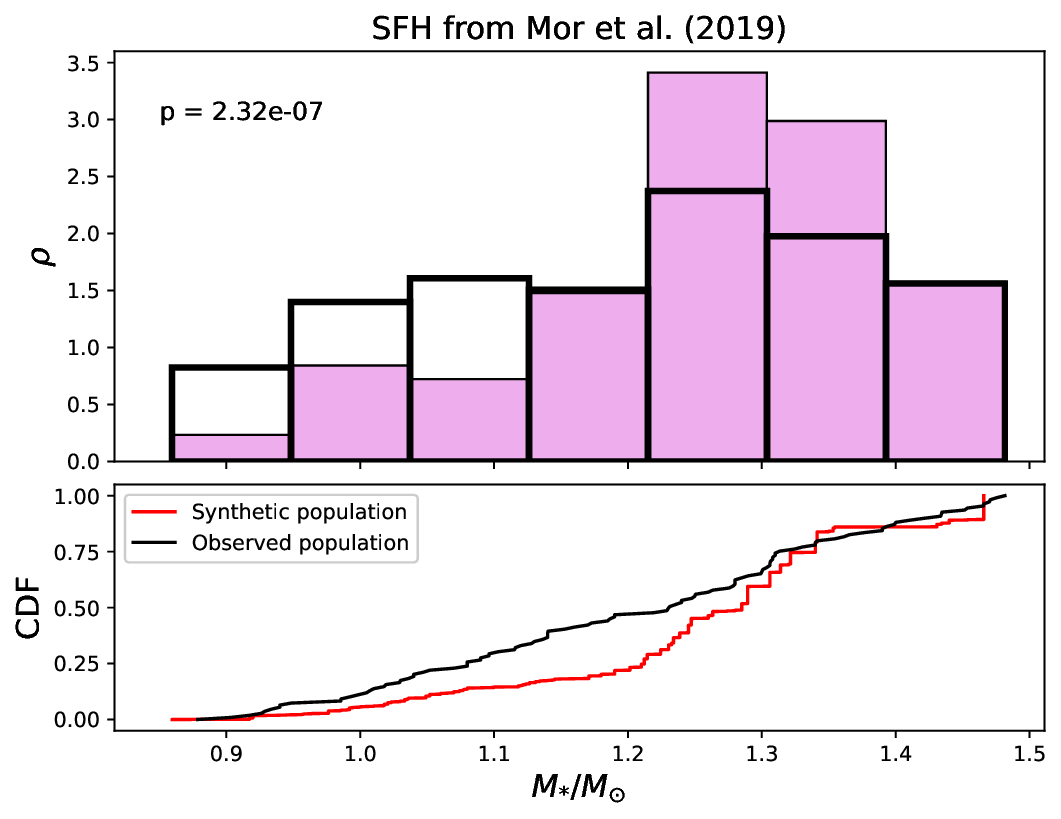}\par
\includegraphics[width=\columnwidth]{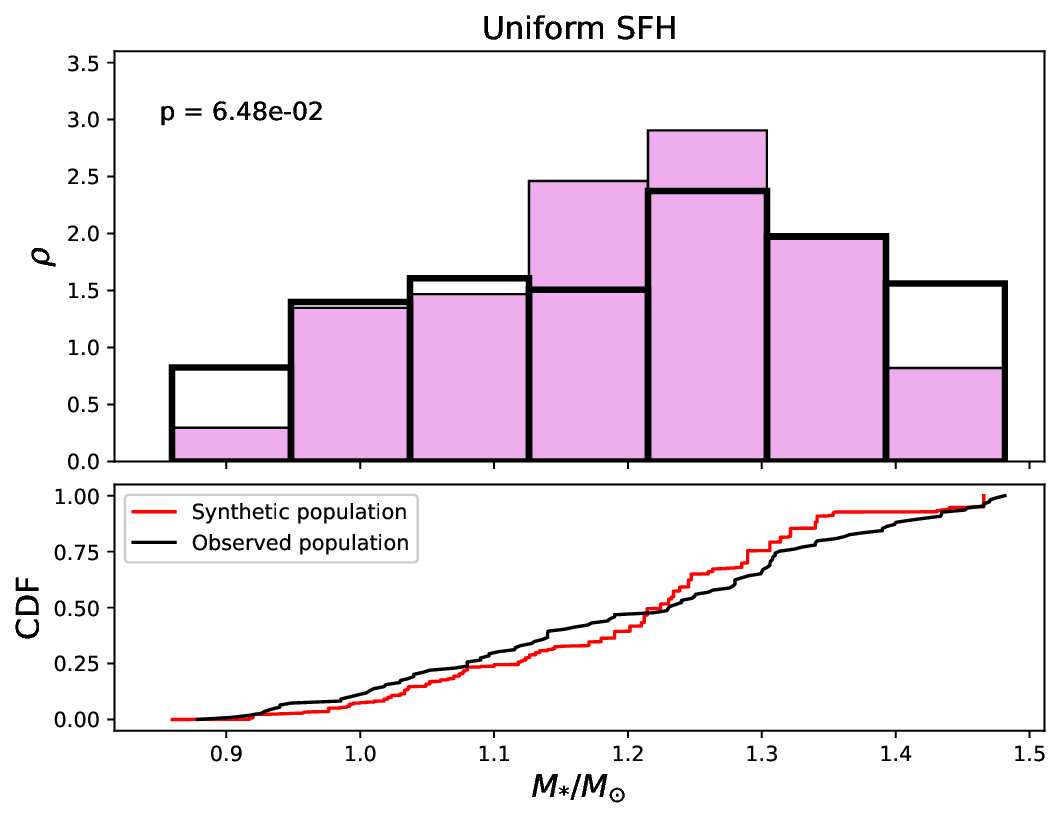}\par
\end{multicols}
    \caption{Probability density functions of the stellar mass for the synthetic populations (purple histogram) and for the observed systems with $t > t_\mathrm{gw}$ (black contour).}
    \label{fig16}
\end{figure*}

A comparison of the stellar mass distributions for the observed and synthetic populations, as presented in Fig.~\ref{fig16}, suggests that the uniform SFH provides a superior fit to the observed distribution. In contrast, the SFH from \cite{Mor2019} underestimates the number of HJs orbiting stars with radiative cores. This discrepancy is primarily due to the star formation burst predicted by \cite{Mor2019} approximately 2-3 Gyr ago. Since this time falls within the MS lifetimes of stars with masses below $1.5 \; M_{\odot}$, the burst leads to an increased number of stars with convective cores in the synthetic population. The obtained $p$-values of the KS tests indicate that the difference between the synthetic population generated with SFH from \cite{Mor2019} and the observed population is statistically significant ($p\ll 0.05$). However, the distribution based on the uniform SFH is relatively close to matching the observed distribution with $p=0.065$.

\begin{figure*}
\begin{multicols}{2}
\includegraphics[width=\columnwidth]{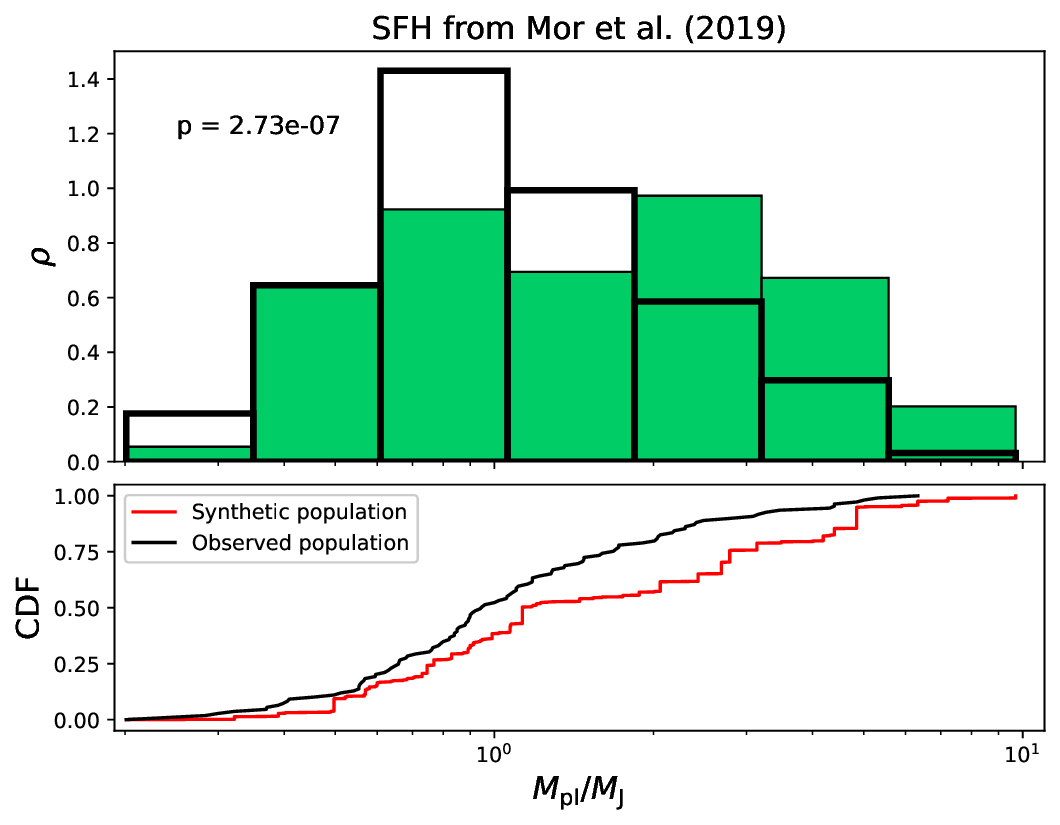}\par
\includegraphics[width=\columnwidth]{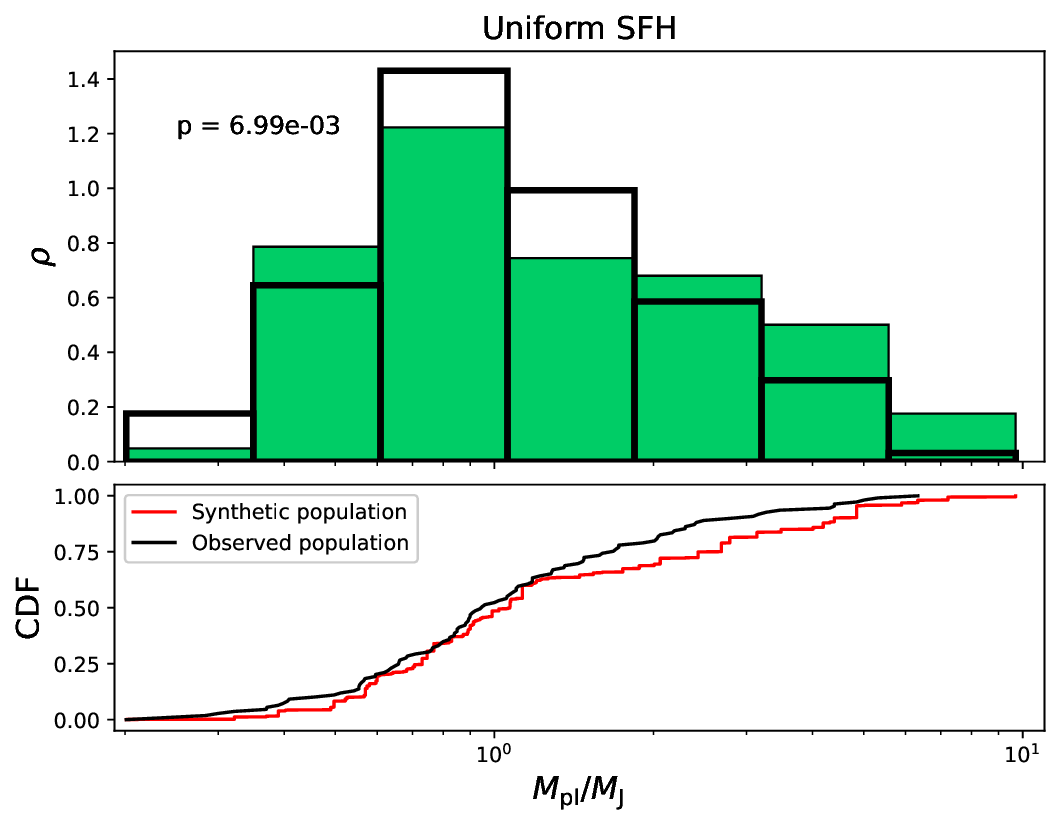}\par
\end{multicols}
    \caption{Probability density functions of the planetary mass for the synthetic populations (green histogram) and for the observed systems with $t > t_\mathrm{gw}$ (black contour).}
    \label{fig17}
\end{figure*}

The same is true for the planetary mass distributions, as demonstrated in Fig.~\ref{fig17}. Again, both synthetic populations diverge from the observed one, with the uniform SFH providing a better fit to the observational statistics.

\begin{figure}	\includegraphics[width=\columnwidth]{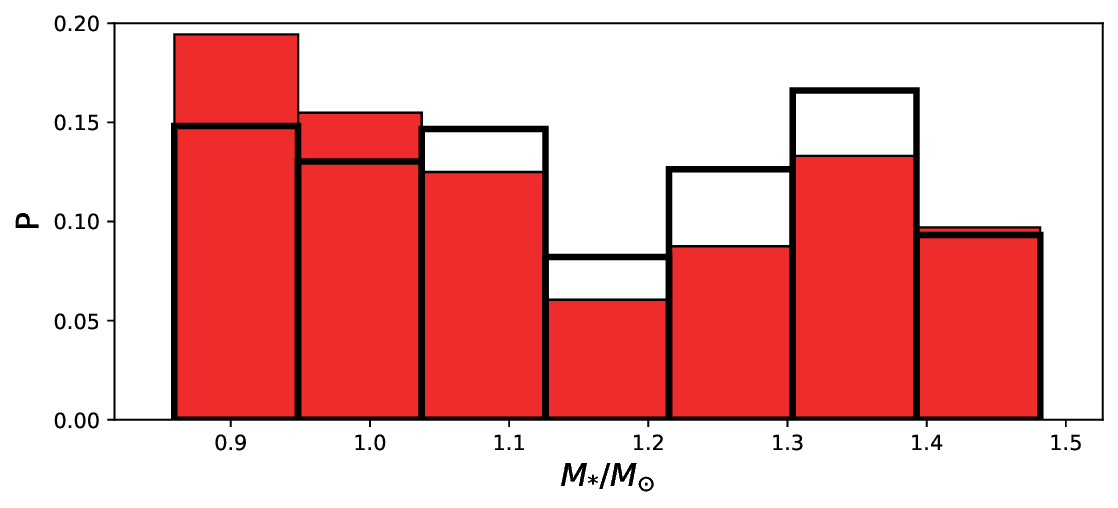}
    \caption{Engulfment probability (defined as the fraction of engulfed HJs prior to the TAMS or 10 Gyr) as a function of stellar mass. Red histogram: SFH from~\protect\cite{Mor2019}. Black contour line: uniform SFH.}
    \label{fig17_1}
\end{figure}

According to our synthetic populations, 12 -- 13\% of stars that once hosted a HJ and have not yet evolved off the MS have engulfed their HJ. The above fraction is calculated relative to all synthetic systems, including young and non-transiting ones. This finding is in good agreement with \cite{L23}, although the latter study focused on solar-mass stars and considered a uniform SFH between 0 and 7 Gyr. The dependence of HJ engulfment probability on stellar mass is presented in Fig.~\ref{fig17_1}. Furthermore, we can estimate the relative number of HJs producing transit timing variations detectable with modern facilities over 10 years of observations. The cumulative shift in transit times $T_\mathrm{shift}$ can be derived from the following \citep[e.g.][]{Wilkins2017}:
\begin{equation}
T_\mathrm{shift} = \frac{27}{8}n \left( \frac{M_\mathrm{pl}}{M_{\star}}\right)\left( \frac{R_\mathrm{*}}{a}\right)^5\frac{1}{Q'}T_\mathrm{dur}^2,
    \label{eq:decay}
\end{equation}
with $T_\mathrm{dur}$ the duration of the observations. We use $T_\mathrm{shift} = 10 \; \rm s$ as a threshold to highlight systems whose orbital decay can be observed within a 10-year period.

Our models predict that, on average, 2.1 -- 2.4 of every 100 migrating transiting HJs exhibit transit timing variations available for detection according to this threshold. This prediction is consistent with current observational data, according to which WASP-12 remains the only star-planet system with orbital decay confirmed by multiple studies. Note that in the observed sample of systems with $P_\mathrm{orb} < 4$ days, there are 110 systems with $t > t_\mathrm{gw}$.

In summary, the excellent agreement between the orbital period distributions of the observed and synthetic planets with $t > t_\mathrm{gw}$ supports the validity of our IGW-driven migration model. However, the lack of agreement for the planetary and stellar mass distributions highlights that there are still limitations of our adopted approach.

\section{Discussion}
\label{sec:discussion}
\subsection{Can resonance locking explain observations?}
In the previous sections, we did not consider g-mode resonances or resonance locking, a potentially important mechanism for planetary migration \citep[e.g.][]{Witte1999,MaFuller2021}. MS stars develop a dense spectrum of gravity modes whose frequencies change with stellar evolution, making it highly likely that a planet's tidal forcing will achieve resonance with at least one of them at some point in the system's lifetime. Subsequently, this resonance can amplify wave amplitudes, potentially leading to effective tidal dissipation. Unless the tidal and mode frequencies evolve together to maintain the resonance, we would expect such a resonance to be passed through quickly, and so its effects on planetary migration may be minor and short-lived. The exceptions are if wave breaking and critical layer formation occur in resonance (in which case we would expect the fully damped regime described by equation~(\ref{eq:tide_gw1}) to subsequently apply, even if equation~(\ref{eq:tide_gw2}) would not predict wave breaking by itself), or if the tidal and mode frequencies can evolve together to maintain a resonance lock.
While a planet is locked in such a resonance, its orbital migration is governed by the evolution of the host star's g-mode eigenfrequencies, potentially allowing for rapid migration \citep{MaFuller2021}. One of the most prominent features of this resonance locking scenario is the positive correlation between the tidal dissipation rate and the orbital period, in contrast to other proposed mechanisms for tidally driven planetary migration (e.g., equation~\ref{eq:tide_gw1} predicts that dissipation scales with an inverse power of the orbital period). This correlation has been proposed to align with the observed distribution of orbital decay timescales for planets \citep{Milholland2025}, as well as the enhanced rotation rates of HJ hosts \citep{Penev2018}. Nevertheless, the resonance locking scenario has been challenged theoretically by \cite{Guo23}, who demonstrated that IGWs can modify the background rotation of a star's central regions through nonlinear feedback, even if wave breaking is not predicted. This effect could significantly alter the conditions required for resonance locking, particularly in stars with radiative cores \citep[see also the discussion in][]{MaFuller2021}.

To investigate the resonance-locking scenario in this work, we utilized a grid of stellar models with a metallicity of [Fe/H]= + 0.2 and masses ranging from 0.8 to 1.1 $M_\mathrm{\odot}$ (specifically, $M_{\star} = 0.8, 0.9, 0.95, 1.0, 1.05$ and $1.1 M_\mathrm{\odot}$). For these models, we employed the GYRE code \citep{GYRE} to compute the non-adiabatic free oscillation modes. We intentionally focus on stars possessing radiative cores, as the feasibility of resonance locking in stars with convective cores is problematic for two reasons. First,  magnetic braking is weaker in higher-mass stars, which would cause planets orbiting them to remain outside their corotation radii for an extended time (where tidal migration would be outward). Second, the g-mode frequencies in stars with convective cores tend to decrease for a significant fraction of their main-sequence lifetimes, which would preclude inward migration due to resonance locking. This is because resonance locking requires the increase in tidal frequency with orbital decay to be balanced by a similar increase in the mode frequencies for a resonance lock to be maintained.

Additionally, we extended our sample of observed planets by increasing the upper limit of the supply region from 5 to 10 days. This is necessary to account for the rapid migration rates predicted by the resonance-locking scenario for planets with longer orbital periods.

\begin{figure*}
\begin{multicols}{2}
\includegraphics[width=\columnwidth]{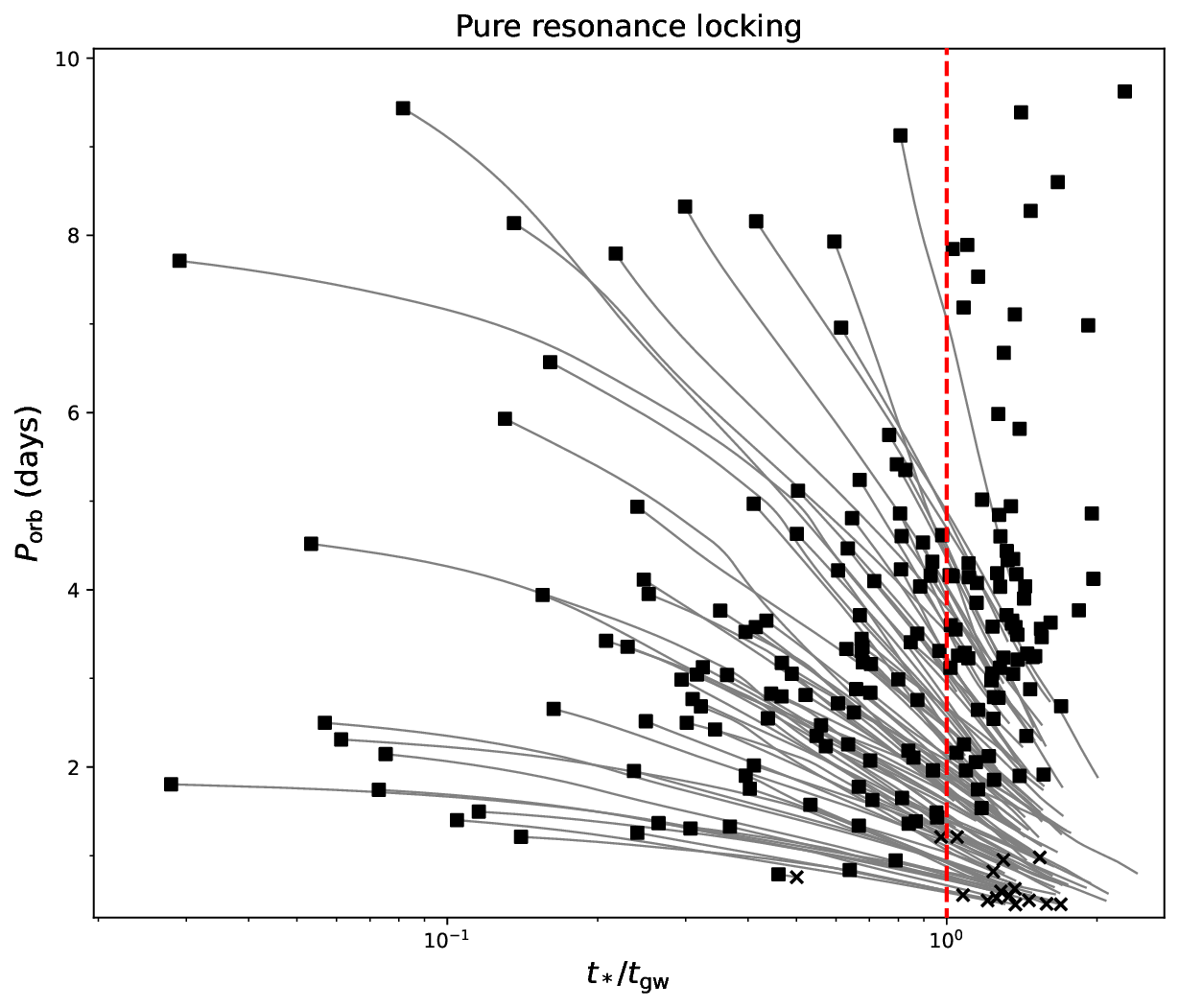}\par
\includegraphics[width=\columnwidth]{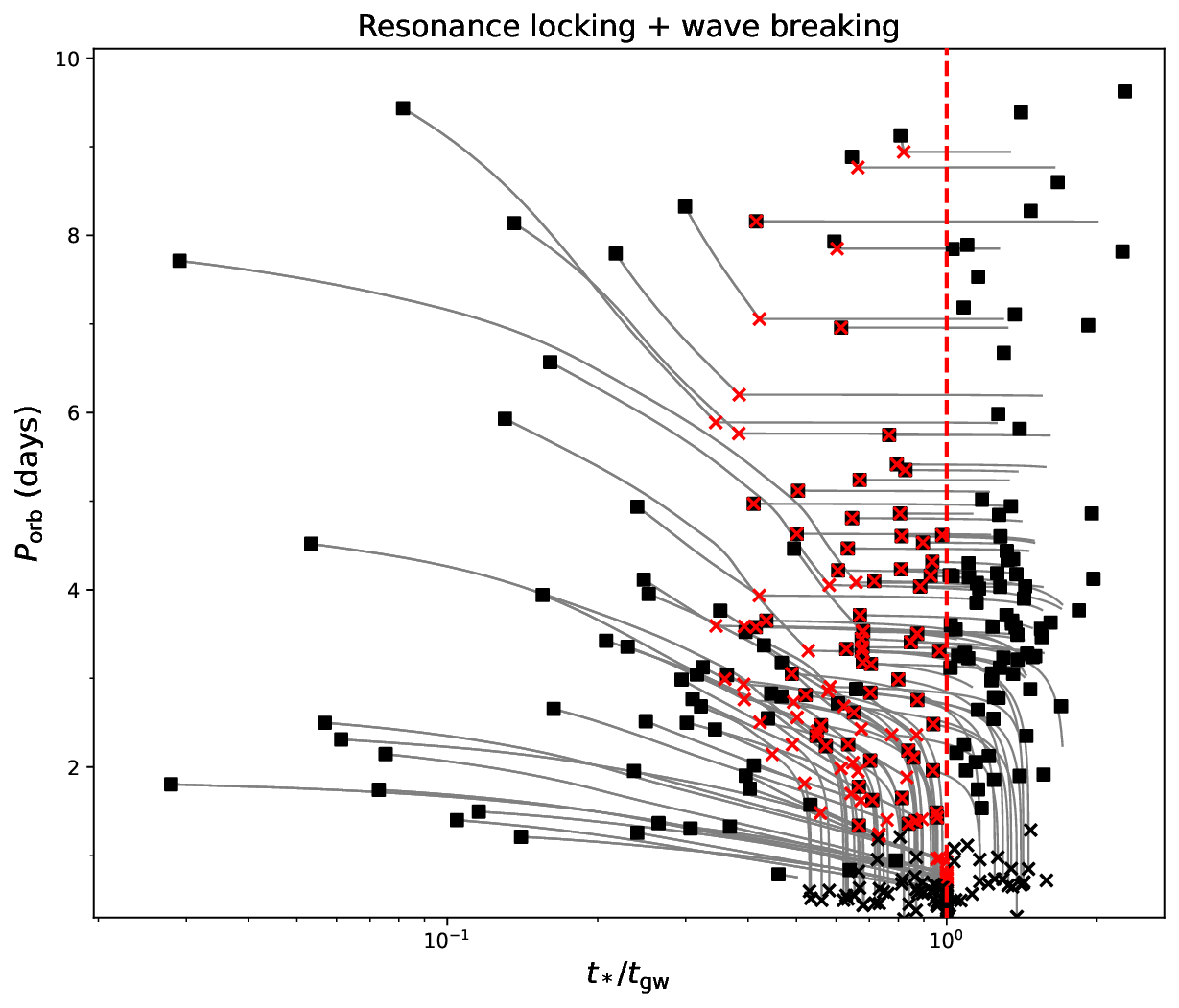}\par
\end{multicols}
    \caption{Distribution of the observed HJ systems around stars with radiative cores in the age–orbital period diagram. The lines show the migration of the planets with $t < t_\mathrm{gw}$. Left panel: pure resonance locking model. Right panel: resonance locking coupled with wave breaking in resonance. Black crosses indicate Roche-lobe overflow; red crosses indicate the onset of wave breaking in resonance.}
    \label{fig18}
\end{figure*}

For every HJ in our sample, we identify the mode whose resonance location is closest to the observed star-planet separation. We then treat this mode as the one in which the planet is trapped in resonance. Following equation (13) from \cite{MaFuller2021} (we note that in their equation (13), $t_\alpha$ needs to be replaced with $t_\mathrm{tide}$), we also neglect the stellar rotation by setting $P_\mathrm{tide} = P_\mathrm{orb}/2$.
In the left panel of Fig.~\ref{fig18} we display the orbital trajectories of planets subject to resonance locking. For the sake of visibility, only the migration of planets with $t_{\star} < t_\mathrm{gw}$ is shown. It is clear that the pure resonance locking scenario inverts the HJ orbital period distribution, resulting in the overabundance of the shortest-period planets, in contrast with the observations. Hence, such a naive application of resonance locking is incompatible with observations.

To align the synthetic population with the observations, wave breaking needs to be taken into account. When a planet is trapped in resonance, non-linear effects are expected to become important earlier than they would have done outside of resonance, i.e., for smaller tidal amplitudes. To model the onset of IGW breaking within a resonance locking scenario, we adopt the prescriptions from \cite{Fuller17} (equations~(13) and (42)-(43)) to determine the mode amplitude. In the right panel of Fig.~\ref{fig18}, which demonstrates the orbital migration according to the combined resonance locking and wave breaking model, the initiation of wave breaking is shown with red crosses. It is important to note that $t_\mathrm{gw}$ still marks the onset of wave breaking if resonance locking is neglected for a given system. For a significant fraction of HJs with $t_{\star} < t_\mathrm{gw}$, whose present-day locations are shown by black squares and overlaid by red crosses, IGW breaking is expected to have taken place if the planet has been trapped in resonance. Specifically, this applies to the planets with $P_\mathrm{orb} \geq 3$ d, as for resonantly locked planets, nonlinear effects increase more strongly with the tidal forcing period.

\begin{figure*}
\begin{multicols}{2}
\includegraphics[width=\columnwidth]{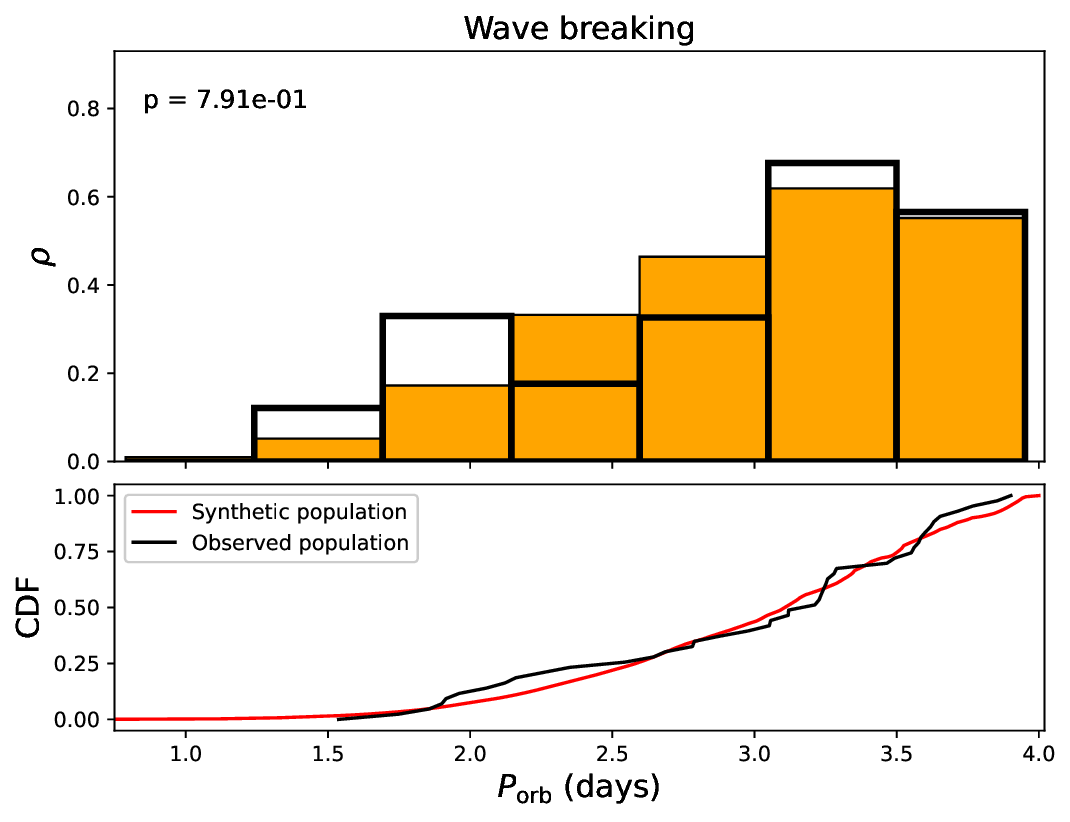}\par
\includegraphics[width=\columnwidth]{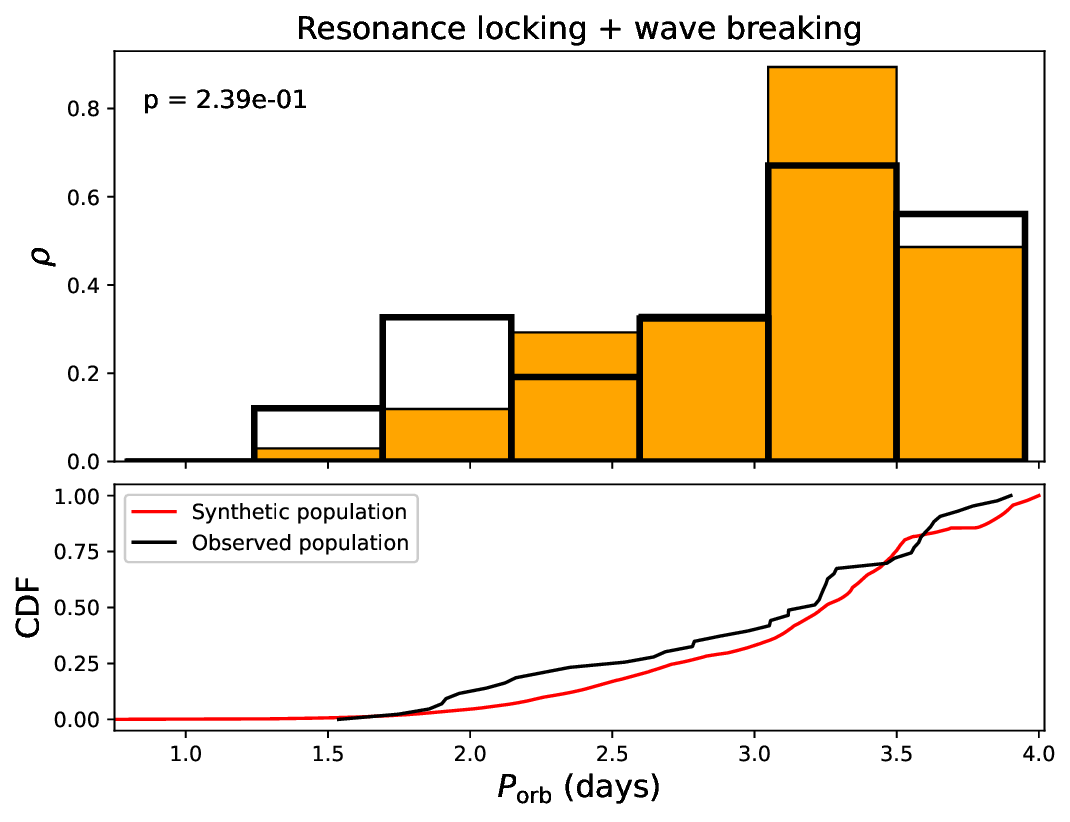}\par
\end{multicols}
    \caption{Same as Fig. \ref{fig15}, but only for systems with $M_{\star} < 1.15 \; M_{\odot}$. HJFH is modelled based on SFH from \protect\cite{Mor2019}.
    Left panel: the synthetic population is generated using an out-of-resonance wave breaking model (equation~\ref{eq:tide_gw2}) Right panel: the synthetic population is generated using a resonance locking model coupled with wave breaking.}
    \label{fig19}
\end{figure*}

Incorporating wave breaking into the resonance locking model restores the convergence between the orbital period distributions of the observed and synthetic populations, as shown in the right panel of Fig.~\ref{fig19}, with the KS test $p-$value exceeding 0.05. Nevertheless, the orbital period distribution of the population generated using the non-resonant wave breaking model, shown in the left panel, appears to align better with the observations.

Our analysis of resonance locking here has a few limitations that should be noted. First, our assumption of zero stellar spin remains valid as long as the planets become trapped in resonance after 1 Gyr; otherwise, neglecting rotation could lead to misidentifying the exact mode a planet is trapped in resonance with. Second, as previously discussed, wave breaking begins earlier for planets with a larger orbital period. Consequently, even if a system's current parameters (orbital period, mass, and age) suggest the resonance-locking criterion should still hold, the planet may have already initiated wave breaking at a larger orbital separation and subsequently migrated inward. A similar effect may occur with wave conversion in stars that possess convective cores, as discussed in detail in Section \ref{subsec:uncert}. Despite these complexities, in this work we consider planets to be trapped in resonance if the system's present-day properties do not indicate that wave breaking should have been induced. Although this approach is an oversimplification, it is not expected to substantially affect the results of our calculations.

In summary, our results suggest that the `pure' application of the resonance locking mechanism — without considering non-linear effects — overestimates the migration rates of HJs, leading to an excess of short-period planets that is not seen in the data. The discrepancy between our conclusions and those of \cite{Milholland2025} likely stems from the limitations associated with fitting a universal scaling law for $Q'$ in their study. We posit that this approach is likely insufficient to capture the complexity of the HJ population; instead, any robust empirical constraint on tidal theory must account for multiple migration regimes. Our findings imply that, while resonance locking may dominate during early stages or at larger separations, it must eventually transition into a wave-breaking regime to remain consistent with the observed orbital period distribution. This transition, particularly relevant during the later stages of the MS lifetime, underscores the necessity of testing future tidal models against both orbital period distributions and measured decay timescales simultaneously.

\subsection{Uncertainties in dissipation mechanisms}
\label{subsec:uncert}

In the present study, we explore a strongly non-linear regime of IGW dissipation in stars with radiative cores, which is governed by the wave-breaking criterion from \cite{Barker} \citep[see also][]{Goodman,OL2007,BO21}. However, as shown in \cite{BO21,Guo23}, at smaller tidal amplitudes, efficient wave dissipation can still occur through a gradual spin-up of the central stellar regions. This process leads to the formation of critical layers, where the fluid's angular velocity matches the orbital frequency of a planet. Close to these layers, IGWs are effectively absorbed, allowing the system to enter a travelling-wave regime \citep[see also][]{Alvan13}. \cite{Guo23} reported that for the small Prandtl numbers (the ratio of kinematic viscosity to thermal diffusivity) expected in stellar interiors, subcritical (non-breaking according to equation~\ref{eq:tide_gw2}) waves can more easily modify the rotational profile of the central stellar region, thus producing a critical layer. In our formalism, this would be equivalent to the critical mass being several orders of magnitude smaller than that predicted by equation~(\ref{eq:tide_gw2}). To understand the importance of this effect, we performed two additional sets of simulations, assuming that $M_\mathrm{crit}$ is reduced by factors of 10 and 100, respectively. We find that this reduction weakens the agreement between our synthetic population and the observed old HJs. Hence, on purely observational grounds, these prescriptions are disfavoured compared with the wave breaking criterion in equation~(\ref{eq:tide_gw2}). This could potentially be explained theoretically by the action of strong, large-scale magnetic fields that can slow down or even prevent the gradual formation of critical layers in this way (compared to the rapid deposition of angular momentum by wave breaking) by damping differential rotation \citep[e.g.][]{MestelWeiss1987,AB25,Skoutnev2025} or potentially via (magneto-)hydrodynamic turbulent transport processes.

Stars with convective cores are known to rotate fast due to less effective magnetic braking compared with stars with radiative cores. Consequently, neglecting stellar rotation when calculating the tidal period may be a significant oversimplification. Prograde rotation tends to decrease the tidal frequencies, which would mean that weaker magnetic fields are required for wave conversion, potentially leading to an earlier onset of IGW dissipation (smaller $t_\mathrm{gw}$). Conversely, for planets orbiting hot stars, which are known to have a wide range of obliquities and a high fraction of retrograde orbits \citep[e.g.][]{Winn2010,Albrecht2012,Knudstrup2024}, assuming zero stellar spin may be a better approximation than considering realistic rotation with aligned orbits. We address the impact of stellar spin in Appendix~\ref{appendix:A}, where we show that accounting for stellar rotation in $P_\mathrm{tide}$ requires a more restrictive IGW conversion criterion to maintain the convergence between the observed HJs and our synthetic population.

In addition, non-circularity and non-zero obliquities will qualitatively modify the picture, as various tidal components are excited beyond the one with $l = m = 2$, each with a different tidal frequency. As obliquity measurements are currently available for only a small sample of planets, this effect cannot be explored further in this work.

When calculating planetary orbital migration around stars with convective cores, we assume that efficient damping of IGWs does not cease even if the radial magnetic field strength subsequently drops below the critical value due to a decreasing tidal period. This is based on assuming that once a ``fully damped regime'' is established, tidal torques will effectively spin up the central regions of the star (outside the convective core) near to where the waves are converted (if they damp nearby). Even without sufficient spin-up to form a critical layer, the resulting partial spin-up would still increase the tidal period in these inner regions relative to a non-rotating scenario, and thus increase the prospects of their subsequent magnetic wave conversion or linear radiative (or Ohmic) damping. Consequently, it is plausible that this self-regulating regime of IGW dissipation could persist under realistic conditions, though we cannot validate this hypothesis because we assume solid-body rotation throughout our work. Further work should explore these aspects.

Due to the dependence of the critical magnetic field strength on the tidal period, using the parameters of the observed orbital period to calculate $t_\mathrm{gw}$ has some limitations. In particular, for several planets initially located close enough to their host stars ($P_\mathrm{orb}$ < 2.5 days), the ratio $t/t_\mathrm{gw}$ can decrease over time as a result of planetary inward migration, reaching values as low as 0.8. Thus, the dynamical state of 7 HJs with $0.8 \leq t/t_\mathrm{gw} \leq 1.0$ and $P_\mathrm{orb}$ < 2.5 days orbiting stars with masses above 1.15 $M_{\odot}$ remains ambiguous. They could be either non-migrating planets that have not yet reached $t/t_\mathrm{gw} = 1.0$ or migrating planets that have reduced their $t/t_\mathrm{gw}$ below unity. These planets are HATS-35 b, HATS-42 b, HD 86081 b, WASP-12 b, WASP-74 b, WASP-92 b and WASP-103 b. Notably, WASP-12 b, for which our prescriptions coupled with the data from \cite{Swastik24} yield, $t/t_\mathrm{gw} = 0.87 \pm 0.03$, remains the only planet with confirmed orbital decay \citep[e.g.][]{Yee2020}, which heightens our concern about the correct identification of planets into migrating and non-migrating populations. 

To clarify which of the HJs mentioned above may be migrating planets that have re-entered the $t/t_\mathrm{gw} < 1$ region, we simulated their orbital evolution backward in time and tested their variation in $t/t_\mathrm{gw}$ as if they were migrating. Our results indicate that only HD 86081, WASP-12, and WASP-103 could have had a higher $t/t_\mathrm{gw}$ in the past, with only HD 86081 b and WASP-103 b reaching $t/t_\mathrm{gw} = 1$. However, for WASP-103 b, IGW-driven migration remains unlikely, as its orbital decay timescale is more than three orders of magnitude lower than the time preceding the onset of IGW dissipation. This makes it statistically more probable to observe this planet in a non-migrating state. The absence of transit-timing variations of WASP-103 b, reported by \cite{Alvarado2024}, confirms this idea. In the case of WASP-12 b, the maximal $t/t_\mathrm{gw}$ is 0.92, which remains below unity even when age uncertainties are considered. Nevertheless, given the uncertainties in IGW conversion and age determination using isochrone fitting, the observed migration of WASP-12 b is consistent with our theoretical framework. The fact that its $t/t_\mathrm{gw}$ value is neither significantly below nor significantly above unity supports this consistency. Finally, for HATS-35 b, HATS-42 b, and WASP-92 b, the upper limits on their ages from observations allow the planets to be older than the time when IGW conversion begins.

If we were to switch off IGW dissipation when $B_r$ equals $B_\mathrm{crit}$, the orbital migration of HJs would be governed by the timescale of evolution of $B_\mathrm{crit}$ rather than the tidally driven decay timescale for a given $Q'$. Consequently, this could lead to a pile-up of migrating HJs with $P_\mathrm{orb} \sim 2$ days and $t/t_\mathrm{gw} \sim 1 $. Due to the limited sample size and uncertainties in age determinations, we can neither confirm nor rule out the presence of this pile-up based on the distribution of observed planets in the age–orbital period diagram. The absence of a detected HJ pile-up in the observed period distribution, as shown in Fig.~\ref{figA1}, tentatively suggests that orbital migration does not slow down when $B_\mathrm{crit}$ reaches $B_r$. However, only a handful of observed systems would experience significant deviations in their expected orbital decay timescales due to this effect, making it likely that its overall impact on the evolution of the HJ population would be negligible.

Finally, we recall that our stellar models do not include element diffusion, which constitutes another significant simplification. As shown in Fig.~\ref{fig1}, this omission leads to an underestimation of the critical magnetic field strength at late stellar ages compared to models that incorporate element diffusion. Consequently, including element diffusion would increase $t_\mathrm{gw}$, particularly for systems where wave conversion is expected to occur toward the end of the MS lifetime, such as systems with $P_\mathrm{orb} \sim 1$ days. Therefore, it would render the observed orbital decay of WASP-12 b inexplicable within the framework of IGW dissipation through wave conversion. When applied to our synthetic populations, accounting for element diffusion would reduce the predicted fraction of HJs that undergo engulfment before TAMS.

\subsection{Other caveats}

The analysis in Section~\ref{sec:population} is based on using a single HJ formation history (HJFH) for every planet in our sample. This approach simplifies the complex problem of planet formation by assuming that the process of HJ formation is uniform and does not depend on the specific stellar or orbital parameters of each system. While three possible scenarios have been proposed to explain their origin -- in-situ formation, primordial disk migration, and high-eccentricity migration \citep[e.g.][]{DawsonJohnson} -- multiple mechanisms likely contribute to the observed population. The relative importance of each pathway is likely influenced by factors such as the final star-planet separation, stellar mass, and metallicity. Given this complexity, adopting a single HJFH may be an oversimplification. The problem of HJ formation remains a subject of ongoing debate. The proposed bimodal nature of the HJ population, supported by the findings of \cite{Nelson2017,HamerSch2022,Wu2023}, has been called into question by \cite{Yee2023}. Their finding of no dependence of the orbital period distribution of close-in Jovian planets on the metallicity of their host stars suggests a single, dominant formation channel across the entire range of parameters considered in the present work.

Another source of uncertainty is the contribution of planets that arrive at their present-day locations via high-eccentricity migration at late ages ($t > t_\mathrm{gw}$), which we refer to as `late-travelers'. Because our approach is designed to show a connection between the populations of young and old HJs, the late-stage implantation of planets would decrease the convergence between our synthetic and observed populations. Nevertheless, neglecting the contribution of these late-travelers is a reasonable simplification, as the dynamical instabilities that drive high-eccentricity migration are expected to take place relatively early in a system's evolution. This is supported by observations of hot Neptunes, which were found to be more abundant in systems with ages between 100 Myr and 1 Gyr \citep{Fernandes2025}. It is possible, though, that a fraction of HJs are formed through secular interactions between planets on longer timescales, as discussed in e.g.~\cite{HamerSch2022,SchSch26}.

Our observed sample of HJs with $P_\mathrm{orb} < 4$ days has a higher fraction of old systems than our models predict. Specifically, 49\% of these planets are old ($t > t_\mathrm{gw}$, which exceeds the predictions of our synthetic models (44\% for the SFH from \citealt{Mor2019} and only 33\% for a uniform SFH). The observed sample's median age of 2.81 Gyr is also higher than the median ages of our synthetic populations, which are 2.69 and 2.72 Gyr, respectively. The primary source of this mismatch is the population of HJs orbiting stars with radiative cores. In our observed sample, which contains 120 such stars, wave breaking is expected in 44 systems (37\%). In contrast, our synthetic models show that these systems are much rarer, comprising only 17\% and 19\% of the populations generated using the SFH from \cite{Mor2019} and a uniform SFH, respectively. Overall, the observed HJs are older compared with the predictions of the Galactic SFH.
Our finding appears to contradict recent studies showing that HJ hosts are kinematically younger and colder than other stars \citep{HamerSch2019, Chen2023, Swastik24, Banerjee2024, Kamulali2025}. This tension can be reconciled by adopting a different SFH model from \cite{Mor2019}, based on the extinction map from \cite{Marshall2006}. The corresponding model (model G12NP-M), which has a more pronounced decreasing trend in SFR at $\tau > 7$ Gyr and a smaller star formation burst at $\tau \sim 3$ Gyr, would lead to a higher median age of synthetic systems (2.84 Gyr) and a higher fraction of HJs engulfed before the TAMS (25\%). Accordingly, the fraction of migrating HJs would rise to 47\%, which is in close agreement with our observed sample. However, the relative number of migrating HJs among systems with stars possessing radiative cores would only increase to 24\%, still significantly below the observed value. Finally, adopting a slightly different version of the SFH will not affect the convergence in the orbital period distribution between the synthetic and observed old populations.

While the ages for the majority of our systems were computed using isochrone fitting based on the homogeneous grid by \citet{Swastik24}, the subsample taken from the NASA Exoplanet Archive includes an admixture of different isochrone models and age-determination techniques. This diversity could potentially introduce systematic biases into our analysis. Nevertheless, typical age uncertainties in our sample likely exceed these possible biases, which is why we expect their impact on our overall conclusions to remain small. Future studies would benefit from repeating this procedure as larger samples of HJ systems with homogeneously determined ages become available.

\subsection{Orbital decay candidates}

Based on our analysis of the observed HJ population, we have identified planets whose estimated ages exceed $t_\mathrm{gw}$ within a 1-$\sigma$ confidence interval. We then filtered this sample to include only those whose orbital migration would result in a cumulative transit timing shift greater than 1 second over a 10-year observation period. These planets are listed in Table~\ref{tab1}.

The most promising systems for upcoming observations are HATS-9 and KELT-14, whose estimated 1-$\sigma$ age ranges exceed $t_\mathrm{gw}$; these are highlighted in green. The planets HATS-9 b and KELT-14 b have been observed for 9 and 10 years, respectively. Our findings underscore the importance of further studying these orbital decay candidates to better constrain IGW breaking and conversion processes. Among the remaining planets, WASP-173 A b, Kepler-17 b, WASP-36 b, and WASP-46 b should be considered prime targets for detecting transit-timing variations. If efficient IGW dissipation is occurring in these systems, their cumulative transit timing shifts are expected to exceed 10 seconds.

\begin{table}
	\centering
    \begin{threeparttable}
	\caption{Planets with the largest predicted transit-timing variations after 10 years of observations}\label{tab1}
	
	\begin{tabular}{|c|c|c|c|c|c|c|} 
		\hline
		Planet & $P_\mathrm{orb}$& $M_\mathrm{pl}$& $M_\mathrm{*}$& $t/t_\mathrm{gw}$& $\tau_\mathrm{a}$&$T_\mathrm{shift}$ \\
        & (days) & ($M_\mathrm{J}$) & ($M_\mathrm{\odot}$) & & (Myr) &(s) \\
		\hline
    TOI-1937 A b & 0.95 & 2.01 & 1.07 & $0.79^{+0.68}_{-0.51}$ & 0.54 & 420 \\
    OGLE-TR-56 b & 1.21 & 3.30 & 1.23 & $0.75^{+0.25}_{-0.33}$ & 1.50 & 151  \\
    WASP-173 A b & 1.39 & 3.47 & 1.09 & $0.87^{+0.47}_{-0.47}$ & 4.66 & 48  \\
    \rowcolor{green!20} \textbf{KELT-14 b} & \textbf{1.71} & \textbf{1.28} & \textbf{1.24} & $\mathbf{1.29^{+0.21}_{-0.21}}$ & \textbf{4.86} & \textbf{46}  \\
    \rowcolor{green!20} \textbf{HATS-9 b} & \textbf{1.92} & \textbf{0.82} & \textbf{1.10} & $\mathbf{1.56^{+0.09}_{-0.09}}$ & \textbf{7.73} & \textbf{29}  \\
    Kepler-17 b & 1.49 & 2.45 & 1.02 & $0.95^{+0.45}_{-0.43}$ & 10.1 & 22  \\
    \rowcolor{gray!20} WASP-36 b & 1.54 & 2.30 & 0.92 & $1.17^{+0.34}_{-0.34}$ & 17.5 & 13  \\
    WASP-46 b & 1.43 & 1.90 & 0.90 & $0.96^{+0.21}_{-0.22}$ & 19.1 & 12  \\
    \rowcolor{gray!20} HATS-15 b & 1.75 & 2.17 & 0.93 & $1.16^{+0.19}_{-0.20}$ & 34.2 & 6.6  \\
    TOI-564 b & 1.65 & 1.46 & 1.00 & $0.81^{+0.52}_{-0.45}$ & 34.8 & 6.5  \\
    \rowcolor{gray!20} CoRoT-18 b & 1.90 & 3.47 & 0.88 & $1.40^{+0.42}_{-0.42}$ & 40.6 & 5.6  \\
    WASP-164 b & 1.78 & 2.13 & 0.95 & $0.67^{+0.39}_{-0.39}$ & 75.2 & 3.0  \\
    \rowcolor{gray!20} Kepler-41 b & 1.86 & 0.57 & 1.13 & $1.24^{+0.46}_{-0.32}$ & 76.0 & 3.0  \\
    \rowcolor{green!20} \textbf{TOI-1181 b} & \textbf{2.10} & \textbf{1.18} & \textbf{1.47} & $\mathbf{2.53^{+0.19}_{-0.19}}$ & \textbf{126} & \textbf{1.8}  \\
    \rowcolor{gray!20} HD 86081 b & 2.14 & 1.48 & 1.21 & $0.89^{+0.21}_{-0.21}$ & 134 & 1.7  \\
    
    \rowcolor{gray!20} HATS-23 b & 2.16 & 1.47 & 1.05 & $1.05^{+0.36}_{-0.37}$ & 157 & 1.4  \\
    \rowcolor{gray!20} HAT-P-53 b & 1.96 & 1.48 & 1.05 & $1.09^{+0.25}_{-0.25}$ & 177 & 1.3  \\
    \rowcolor{green!20} \textbf{HAT-P-65 b} & \textbf{2.60} & \textbf{0.55} & \textbf{1.19} & $\mathbf{1.60^{+0.15}_{-0.49}}$ & \textbf{202} & \textbf{1.1}  \\
    \rowcolor{green!20} \textbf{TOI-2977 b} & \textbf{2.35} & \textbf{1.68} & \textbf{0.94} & $\mathbf{1.44^{+0.39}_{-0.44}}$ & \textbf{202} & \textbf{1.1}  \\
		\hline
	\end{tabular}
\begin{tablenotes}
\item The rows highlighted in green indicate the planets where the estimated 1-$\sigma$ age range exceeds $t_\mathrm{gw}$. The rows highlighted in grey indicate the planets whose estimated age exceeds $t_\mathrm{gw}$. The planet HD 86081 b is a notable exception, as its $t/t_\mathrm{gw}$ value is able to decrease due to planetary inward migration (see Section \ref{subsec:uncert}).

\end{tablenotes}
\end{threeparttable}
\end{table}
\section{Conclusions}

\label{sec:conclusions} 

The effects of stellar internal gravity waves (IGWs) on the tidal evolution of star-planet systems have been incompletely explored to date. However, their influence on planetary orbital migration can be crucial, particularly for older stars in the latter half of their main-sequence evolution, for which the HJ occurrence rate was found to decrease with age according to \cite{Miyazaki2023}. In this work, we apply prescriptions for tidal dissipation due to internal gravity waves in stellar radiative zones of approximately solar mass stars with either radiative (stellar types G and K) or convective cores (type F), to investigate the implications of IGW dissipation for planetary systems. We focused on two mechanisms: wave breaking, which is relevant for stars with radiative cores \citep{BO21,Barker}, and conversion to magnetic waves in stars with convective cores \citep{Duguid24}.

Using simple (but physically-motivated) prescriptions for these mechanisms, we have managed to explain the enhanced rotation of both TOI-2458 and GJ 504 by the previous engulfment of a hypothetical HJ driven by IGW dissipation. For TOI-2458, a $\sim 1.05 \: M_{\odot}$ star with a radiative core, we accounted for the presence of its hot Neptune, TOI-2458 b, to constrain the orbital period of the engulfed HJ. Our simulations invoking IGW breaking successfully reproduced the age and rotation period reported by \citet{Subjak} provided that the engulfed planet has a mass exceeding 0.3 $M_\mathrm{J}$ and an orbital period below 2.5 days. In the case of GJ 504, a $ \sim 1.22 \: M_\mathrm{\odot}$ star with a convective core, the parameters of the engulfed planet strongly depend on stellar metallicity. For a metal-poor model with Z = 0.018, our simulations invoking magnetic wave conversion reproduce the observed characteristics of GJ 504 following its coalescence with a planet exceeding 0.66 $M_\mathrm{J}$ with an initial orbital period between 1.5 and 2.25 days. For a metal-rich model with Z = 0.022, the minimum required planetary mass increases to 0.95 $M_\mathrm{J}$, and its initial orbital period range changes to 1.65 -- 2.75 days.

To investigate the impact of IGW dissipation on the planetary population as a whole, we examined a sample of HJ systems with age estimates from \cite{Swastik24}. This sample was supplemented by systems with ages taken from the NASA Exoplanet Archive. Based on our analysis, we separated our sample into two groups: systems that are too young to facilitate efficient IGW dissipation, and old systems where efficient wave dissipation is active. These two subsamples show qualitatively different orbital period distributions. Young systems are characterized by a uniform distribution, while old systems exhibit a steep decline at short orbital periods, which is indicative of effective tidally-driven orbital decay.

To explore the connection between young and old HJs, we conducted a population synthesis study where tidal migration is driven by IGW dissipation. We found that the distribution of old planets can be recreated based on the distribution of young ones, which are treated as the initial population. Our analysis has demonstrated:
\begin{itemize}
    \item Our synthetic populations of old HJs generated using the uniform HJ Formation History (HJFH) and the HJFH based on the fiducial Star Formation History (SFH) from \cite{Mor2019} successfully reproduce the orbital period distribution of the observed sample.
    \item A uniform HJFH yields a reasonable agreement between the stellar and planetary mass distributions of our synthetic and observed populations. However, HJFH from \cite{Mor2019} underestimates the number of HJs orbiting stars with radiative cores.
    \item Both synthetic populations underestimate the median age of HJ systems and the fraction of non-migrating HJs, particularly in systems with low-mass host stars. This discrepancy is partially resolved in a third synthetic population generated using a SFH model G12NP-M from \cite{Mor2019}.
    \item Up to 13\% of stars that once hosted a HJ and have not yet evolved off the MS have engulfed their planet.
    \item Out of every 100 HJ systems with active IGW dissipation and an orbital period below 4 days, 2.1--2.4  systems are expected to exhibit a cumulative transit timing shift exceeding 10 seconds within 10 years of observations.
\end{itemize}

Finally, to guide future observations, we have compiled a list of the most promising candidates for the detection of tidally-driven orbital decay. According to our estimates, HATS-9 b and KELT-14 b are likely to exhibit significant transit-timing variations. This is because their ages, within the 1-$\sigma$ range, are above the age for onset of efficient IGW dissipation. Moreover, their cumulative transit timing shifts over a 10-year baseline exceed 10 seconds, making them ideal targets for upcoming observations. Other promising candidates for future observations include WASP-173 A b, Kepler-17 b, WASP-36 b, and WASP-46 b. These planets have a non-zero probability of undergoing inward migration with an orbital decay timescale sufficiently short to potentially be detected.

\section*{Acknowledgments}

YL was supported by the Foundation for the Advancement of Theoretical Physics and Mathematics BASIS. AJB was supported by STFC grants ST/W000873/1 and UKRI1179.

We gratefully acknowledge Dr. Chowbay Swastik for the provided data on ages and other parameters of star-planet systems. YL thanks Anastasiya Yarovova for helpful discussions. We would also like the thank the referee for their careful reading and constructive suggestions. This research has used the NASA Exoplanet Archive, which is operated by the California Institute of Technology, under contract with the National Aeronautics and Space Administration under the Exoplanet Exploration Program. 
\section*{Data Availability}
The data underlying this article will be shared on reasonable request to the corresponding author.


\bibliographystyle{mnras}
\bibliography{IGW} 



\appendix
\section{Stellar rotation and wave conversion in stars with convective cores}
\label{appendix:A}
In Section \ref{sec:population}, our calculations of $t_\mathrm{gw}$ neglected stellar rotation, which implies $P_\mathrm{tide} = P_\mathrm{orb}/2$. This simplification primarily affects stars with convective cores, as they are expected to sustain rapid rotation for longer periods of their evolution. Furthermore, the minimal magnetic field strength (given by equation \ref{eq:magn2}) scales with the square of the tidal forcing frequency, $\omega_\mathrm{tide}$. Therefore, any reduction in $\omega_\mathrm{tide}$, such as when the planetary orbital period and the star's rotational period are comparable, may facilitate an earlier onset of wave conversion.

\begin{figure*}
\begin{multicols}{2}
\includegraphics[width=\columnwidth]{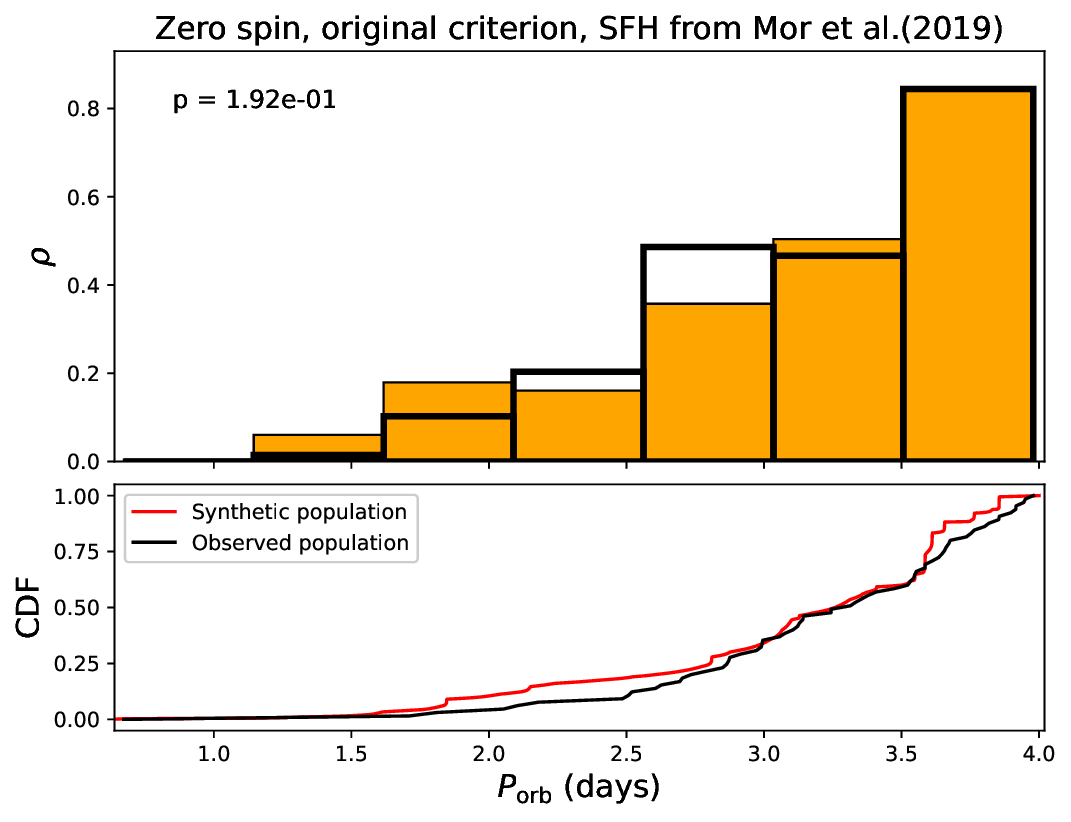}\par
\includegraphics[width=\columnwidth]{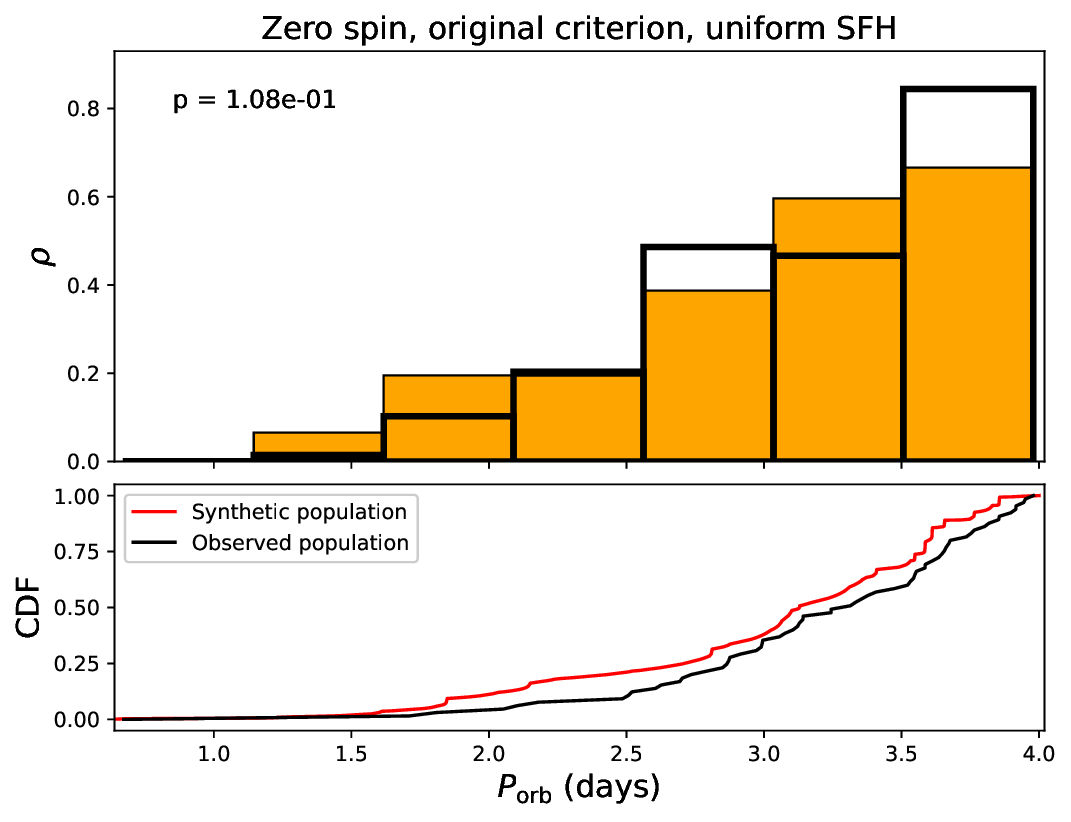}\par
\end{multicols}
    \caption{Same as Fig. \ref{fig15}, but only for systems with $M_{\star} \geq 1.15 \; M_{\odot}$.}
    \label{figA1}
\end{figure*}

Fig.~\ref{figA1} compares the orbital period distributions of the observed and synthetic samples of HJs around stars with $M_{\star} \geq 1.15 M_{\odot}$ and $t > t_\mathrm{gw}$. As in Section ~\ref{sec:population}, stellar rotation is neglected. When the SFH is taken from \cite{Mor2019}, there is good agreement between the observed and synthetic distributions, as confirmed by a high KS $p$-value (greater than 0.05).

\begin{figure*}
\begin{multicols}{2}
\includegraphics[width=\columnwidth]{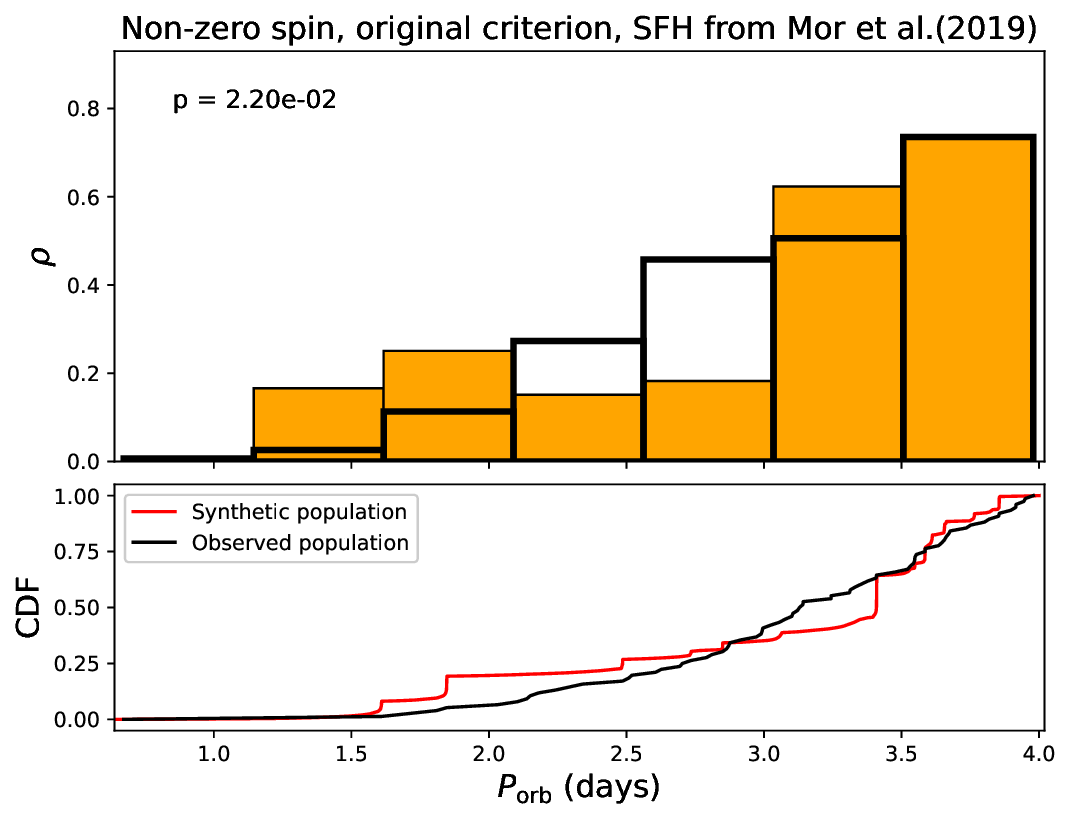}\par
\includegraphics[width=\columnwidth]{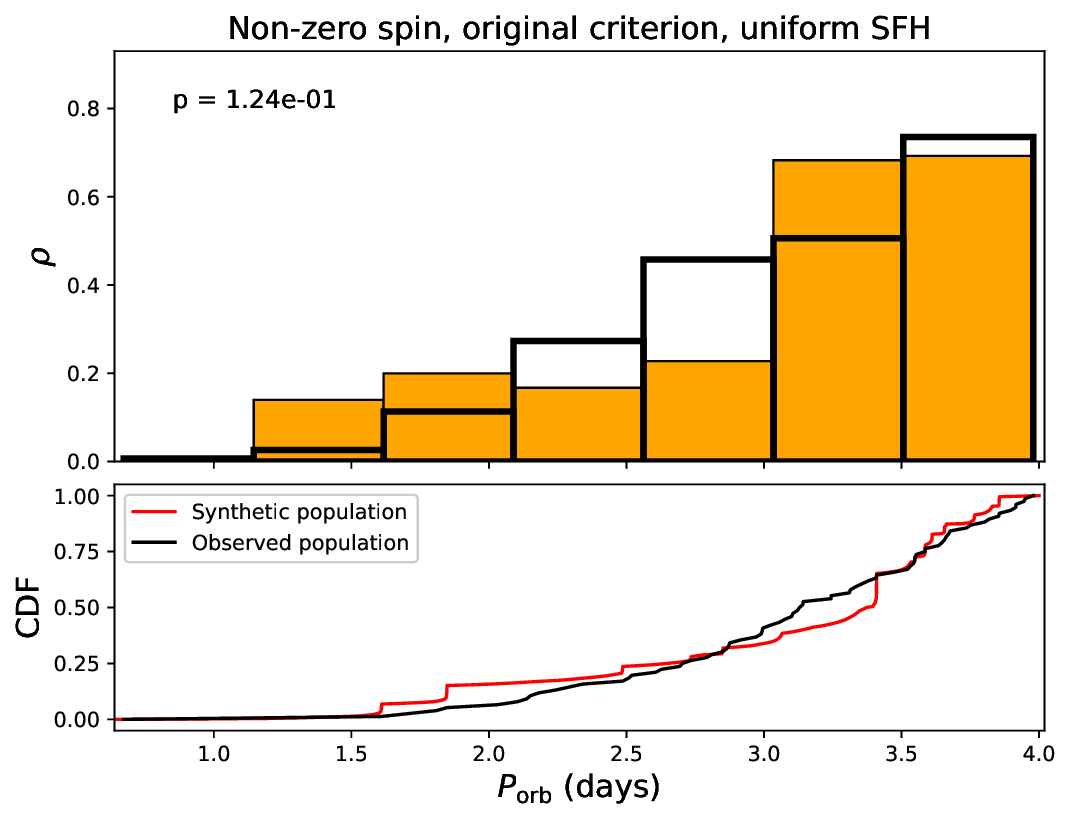}\par
\end{multicols}
    \caption{Same as Fig. \ref{figA1}, but assuming non-zero stellar spin when calculating $P_\mathrm{tide}$.}
    \label{figA2}
\end{figure*}

Next, we generate another pair of synthetic populations, this time taking into account non-zero stellar spin and assuming that the orbits of HJs are all aligned and prograde. Fig.~\ref{figA2} shows that including rotation reduces the agreement between the synthetic distribution based on the SFH from \cite{Mor2019} and the observed distribution, causing the $p-$value to fall below the significance threshold of 0.05. On the other hand, better agreement between the synthetic distribution based on the uniform SFH is observed in the right panel of Fig.~\ref{figA2}.

To improve the agreement when incorporating non-zero stellar spin, we explore imposing a more stringent wave conversion criterion. Specifically, we require the condition $|B_r| \geq B_\mathrm{crit}$ to be satisfied over a radial extent greater than the IGW radial wavelength $\lambda_r$ for magnetic wave conversion to occur. The wavelength derived from the WKB dispersion relation for IGWs is given by:
\begin{equation}
\lambda_r = \frac{2\pi r}{  \sqrt{l(l+1)\left(\frac{N^2}{\omega_\mathrm{tide}^2} -1\right)}},
\label{eq:wavelength}
\end{equation}
with $l = 2$. For our modified wave conversion criterion, we take the minimal value of $\lambda_r$ over the region where $|B_r| \geq B_\mathrm{crit}$ to provide a more conservative estimate. It is unclear which condition is most appropriate, although it would be important to explore theoretically in future work \citep[building upon, e.g.,][]{Lecoanet17,RuiFuller23}.

\begin{figure*}
\begin{multicols}{2}
\includegraphics[width=\columnwidth]{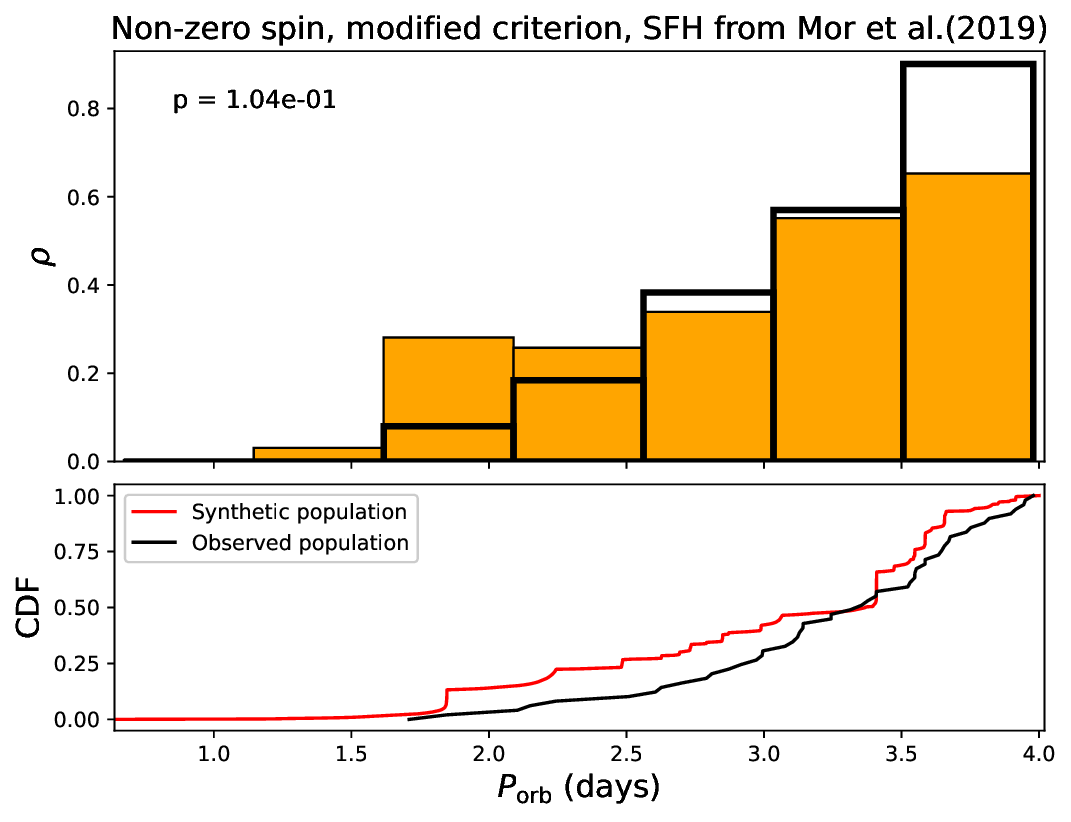}\par
\includegraphics[width=\columnwidth]{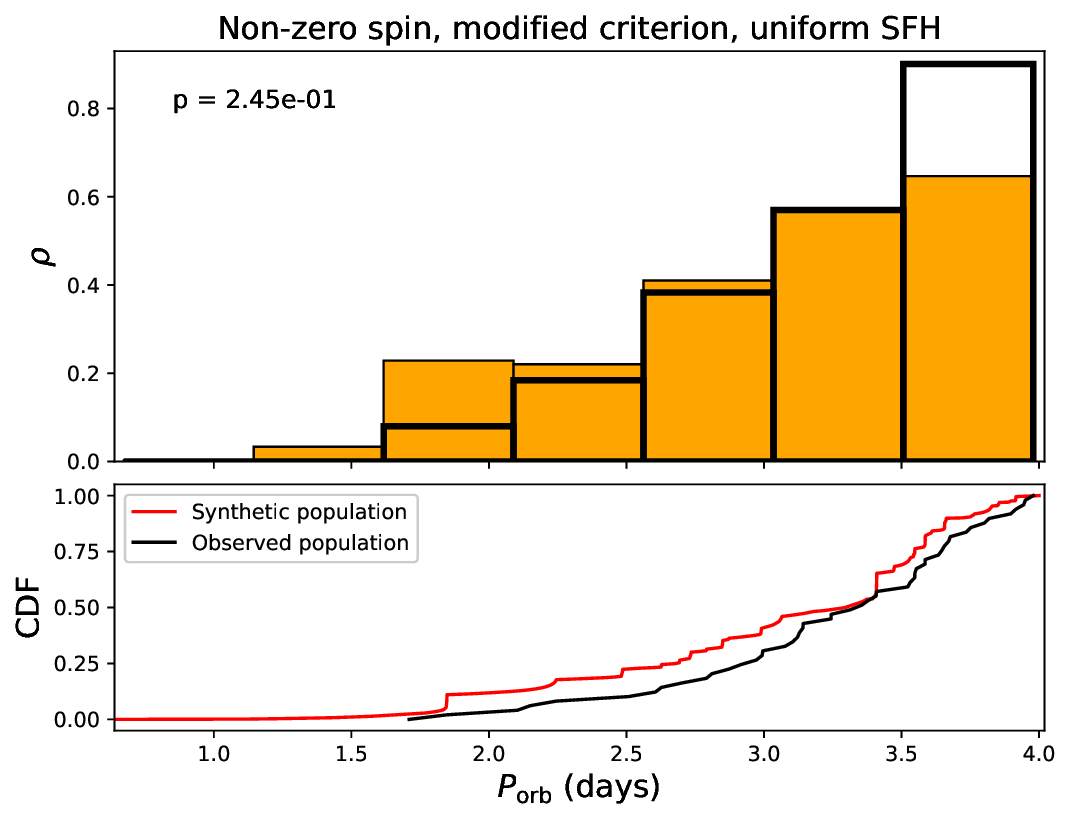}\par
\end{multicols}
    \caption{Same as Fig.~\ref{figA2}, but the onset of wave conversion is determined based on the modified criterion.}
    \label{figA3}
\end{figure*}

Fig.~\ref{figA3} shows that incorporating this modified wave conversion criterion together with stellar rotation increases the KS $p$-value. The synthetic population based on the uniform SFH provides a better match to the observed sample, although in both synthetic populations, HJs with the shortest orbital periods ($P_\mathrm{orb} < 2$ days) are overrepresented.

Overall, the limited number of hot Jupiters orbiting stars with convective cores reduces the statistical power of these tests, preventing us from drawing definitive conclusions about the impact of stellar rotation on the agreement between observed and synthetic samples. Our results tentatively suggest, though, that the incorporation of stellar spin effects should be accompanied by the adoption of a more stringent magnetic wave conversion criterion than the one proposed by \citep{Duguid24}.

\bsp	
\label{lastpage}
\end{document}